\documentclass[fleqn,usenatbib]{mnras}

\usepackage[T1]{fontenc}
\usepackage{ae,aecompl}
\usepackage{natbib,twoopt}
\usepackage{subcaption}
\usepackage{graphicx}	
\usepackage{amsmath}	
\usepackage{amssymb}	
\usepackage{amsfonts}
\usepackage{wasysym}
\usepackage{pslatex}
\usepackage{pdflscape}
\usepackage{comment}
\usepackage{float}
\usepackage{csquotes}

\title[Investigation of Rocket Effect in BRC 18]{Investigation of Rocket Effect in BRC 18 using \textit{Gaia} EDR3}

\author[Piyali Saha et al.]{
Piyali Saha,$^{1,2,3}$\thanks{E-mail: s.piyali16@gmail.com (PS)}
Maheswar G.,$^{1}$
D. K. Ojha$^{4}$
and Sharma Neha$^{5}$
\\
$^{1}$Indian Institute of Astrophysics (IIA), Sarjapur Road, Koramangala, Bangalore 560034, India\\
$^{2}$Satyendra Nath Bose National Centre for Basic Sciences (SNBNCBS), Salt Lake, Kolkata-700 106, India\\
$^{3}$Pt. Ravishankar Shukla University, Amanaka G.E. Road, Raipur, Chhatisgarh 492010, India\\
$^{4}$Tata Institute of Fundamental Research (TIFR), Homi Bhabha Road, Mumbai 400005, India\\
$^{5}$Finnish Centre for Astronomy with ESO (FINCA), FI-20014, University of Turku, Finland
}

\date{Accepted XXX. Received YYY; in original form ZZZ}

\pubyear{2021}

\begin{document}
\label{firstpage}
\pagerange{\pageref{firstpage}--\pageref{lastpage}}
\maketitle

\begin{abstract}
    Bright-rimmed clouds (BRCs) are ideal candidates to study radiation-driven implosion mode of star formation as they are potential sites of triggered star formation, located at the edges of H{\sc ii} regions, showing evidence of ongoing star formation processes. BRC 18 is located towards the eastern edge of relatively closer ($\sim$400 pc) H{\sc ii} region excited by $\lambda$ Ori. We made R-band polarimetric observations of 17 candidate young stellar objects (YSOs) located towards BRC 18 to investigate any preferred orientation of the discs with respect to the ambient magnetic field and the direction of energetic photons from $\lambda$ Ori. We found that the discs are oriented randomly with respect to the projected magnetic field. Using distances and proper motions from the \textit{Gaia} EDR3 of the candidate YSOs, we investigated the possible acceleration of BRC 18, away from $\lambda$ Ori due to the well known ``Rocket Effect'', by assuming that both the candidate YSOs and BRC 18 are kinematically coupled. The relative proper motions of the candidate YSOs are found to show a trend of moving away from $\lambda$ Ori. We computed the offset between the angle of the direction of the ionization front and the relative proper motion of the candidate YSOs and found it to lie close to being parallel to each other. Additionally, we found 12 sources that are comoving with the known candidate YSOs towards BRC 18. These comoving sources are most likely to be young and are missed in previous surveys conducted to identify potential YSOs of the region.
\end{abstract}



\section{Introduction}

The effect of stellar feedback on parameters such as star formation rates, efficiency, and the initial mass function is one of the main focuses of the study of star formation and galaxy evolution. The stellar feedback is considered as positive when it triggers collapse and initiates star formation in otherwise stable molecular condensations and as negative when it hinders star formation by dispersing material, which otherwise would have formed stars. Massive OB stars emit a copious amount of ionizing radiation, which quickly ionizes and heats the surrounding medium. The ionizing heating of the massive OB stars-facing side of pre-existing density structures drives high-pressure waves into them, leading to their compression and subsequent star formation. This mechanism of radiation-triggered star formation is called radiation-driven implosion  \citep[RDI;][]{1989ApJ...346..735B, 1990ApJ...354..529B, 1994A&A...289..559L, 2009ApJ...692..382M, 2012MNRAS.426..203H}. The consequent clouds are called bright-rimmed clouds \citep[BRCs; ][]{1991ApJS...77...59S, 1994ApJS...92..163S}, which are usually found to be located at the edges of H{\sc ii} regions, having a bright rim facing the ionizing source/s and often followed by an extended structure on the opposite side. 

In the RDI mode, the evolution of a dense (optically thick) clump being photo-ionized proceeds in two phases - the collapse and transient phase \citep{1989ApJ...346..735B, 1994A&A...289..559L} and the cometary phase \citep{1990ApJ...354..529B, 1994A&A...289..559L}. The collapse phase, which lasts for $\sim10^{5}$ yr, begins when an ionization front preceded by a shock wave propagates through the cloud, causing it to compress on its axis and forming a high-density core, which eventually could be the potential site for triggered star formation. At the ionization front, the material gets ionized and escapes towards the ionizing source, causing a photo-evaporation flow. The escaping material has an equal and opposite reaction on the cloud, causing it to accelerate away from the ionizing star, similar to a jet emitted from a rocket. This phenomenon, called the ``Rocket Effect'', was proposed by \cite{1954BAN....12..177O}. In the cometary phase, the cloud enters into a quasi-equilibrium state, continues to accelerate away from the ionizing source, and may eventually get photo-evaporated completely. Typically, a B2 star can push a cloud with a speed of $\sim19$ km s$^{-1}$, which eventually goes down to a final speed of $3-4$ km s$^{-1}$ for a typical cloud having mass of $100$ M$_{\odot}$ \citep{1954BAN....12..187K}. The time (t$_{\star}$) at which the star formation starts and the fraction of the cloud material that gets converted into stars critically depends on the amount of incident ionizing flux \citep{2011ApJ...736..142B}. For a typical value of the incident ionizing flux, 10$^{9}$ cm$^{-2}$s$^{-1}$ \citep{2004A&A...415..627T,2004A&A...426..535M}, the t$_{\star}$ is $\sim10^{5}$ yr. Therefore, if the star formation is triggered in BRCs, the young stellar objects (YSOs) that are formed inside the accelerating cloud should share a similar motion and hence would also move approximately radially away from the ionizing star or stars, which is or are responsible for the triggering \citep{2015MNRAS.450.1199D}.

Based on their association with \textit{IRAS} (Infrared Astronomical Satellite) sources, a total of 89 BRCs (44 in the northern hemisphere and 45 in the southern hemisphere) were cataloged by \cite{1991ApJS...77...59S} and \cite{1994ApJS...92..163S}. Signposts of recent and ongoing star formation, e.g., H$\alpha$ emitting stars, Herbig-Haro objects, \textit{IRAS} point sources, molecular outflows, etc., have been noted in BRCs \citep[e.g., Table 1 of][]{1998ASPC..148..150E}. BRCs are, thus, potential sites of radiation-triggered star formation \citep[e.g., ][]{2011EAS....51...45E}. Subsequent studies of BRCs revealed the presence of small aggregates or clusters of YSOs, with the older sources lying closer to the ionizing source and the younger ones lying closer to the \textit{IRAS} source located inside the BRCs. This led to the hypothesis of ``small-scale sequential star formation" \citep[SSSSF;][]{1995ApJ...455L..39S, 2006PASJ...58L..29M, 2007ApJ...654..316G, 2008AJ....135.2323I, 2009MNRAS.396..964C, 2009ApJ...699.1454G, 2010ApJ...717.1067C}.

Magnetic fields are believed to play a significant role in the evolution of molecular clouds as well as in regulating the star formation in them \citep[e.g., ][]{1999ApJ...520..706C, 2014ApJ...785...69C, 2019FrASS...6....5H}. Stars are born as a consequence of the gravitational contraction of molecular cores. During this process, gravity affects neutral species of the cloud while ions are attached to the magnetic field. In a system of clouds with straight magnetic field lines, \cite{1993ApJ...417..243G} showed that the major axis of the pseudo-disc is oriented perpendicular to the magnetic field lines, which is expected if the gravitational collapse happens preferentially along the field lines. Therefore, the orientation of YSOs with circumstellar discs can provide information on the impact of the magnetic field during the gravitational collapse. The circumstellar discs are expected to be aligned with their symmetry axis almost parallel to the cloud's magnetic field \citep{1993ApJ...417..220G}. Now, if the collapse is not significantly affected by the magnetic field, the orientation of the circumstellar discs could be random \citep{2004A&A...425..973M}. 

The presence of circumstellar material around YSOs associated with the BRCs has been established by the observations of strong infrared (IR) excess \citep[e.g.][]{2008AJ....135.2323I, 2013AJ....145...15R, 2014MNRAS.443.1614P, 2016AJ....151..126S} and emission line features \citep[e.g.][]{2002AJ....123.2597O,2008AJ....135.2323I, 2020arXiv200600219H}. The dust grains and electrons present in these flattened discs scatter the light off from the central source and hence, are mainly responsible for the polarized light measured in them \citep{1977IAUS...75..179S, 1978A&A....70L...3E, 1979ApJ...229L.137B, 1982A&AS...48..153B, 1985ApJS...59..277B, 1988PThPS..96...37S}. The degree of polarization depends on the number of dust grains, also on the degree of flatness of the disc and its alignment with respect to the line of sight. For an optically thin disc, if linear polarization is made by single scattering, the position angle of the polarized light is oriented perpendicular to the disc, while in case of an optically thick disc, where the polarization is processed by the scattering dust grains oriented perpendicular to the disc, the position angle is aligned parallel to the disc plane \citep[e.g.][]{1977A&A....57..141B,1978A&A....70L...3E}. Therefore, polarimetry can be used as a tool to obtain the geometrical structure of the circumstellar material around the YSOs \citep{2005ASPC..343..128M}.

In this paper, we present the results of our R-band polarimetric measurements of 17 candidate YSOs identified in the direction of BRC 18 to infer their projected disc orientations. We intend to investigate the impact of the cloud's magnetic field as well as the impact of energetic radiation from the ionizing source, $\lambda$ Ori, on the orientation of the circumstellar discs around the candidate YSOs associated with the cloud. We selected BRC 18 because it is one of the closest BRCs to the Sun \citep[$\sim 400$ pc, using \textit{Gaia} data release 2 (DR2);][]{2018AJ....156...84K} and because of the knowledge of the cloud's magnetic field inferred from the R-band polarization measurements of the stars lying behind it (Neha et al. under preparation; hereafter, Neha et al.).

The distance and the proper motion values obtained from the \textit{Gaia} early data release 3 (EDR3) \citep{2020arXiv201201533G} are utilized to establish the association of the candidate YSOs with BRC 18. Based on optical photometry and multiobject spectroscopy, \cite{2001AJ....121.2124D} made a detailed study towards $\lambda$ Ori H{\sc ii} region, where BRC 18 is located at the eastern edge. Using wide-field near-IR imaging observations, \cite{2012PASJ...64...96H} identified a few candidate YSOs towards multiple BRCs, including BRC 18. \cite{2015AJ....150..100K} conducted optical spectroscopic observations of a number of candidate YSOs identified based on the \textit{WISE} (Wide-field Infrared Survey Explorer) data in the $\lambda$ Ori H{\sc ii} region. \cite{2018AJ....156...84K} used spectroscopic and astrometric data from APOGEE (Apache Point Observatory Galactic Evolution Experiment)-2 and \textit{Gaia} DR2, respectively, to develop six-dimensional structures towards the $\lambda$ Ori H{\sc ii} region, and identified distinct groups of YSOs through a hierarchical clustering algorithm. We also obtained a number of pre-main-sequence (PMS) stars from \cite{2018A&A...620A.172Z}, who made a 3-dimension (D) mapping of young stars in the solar neighbourhood using the \textit{Gaia} DR2 measurements. Recently, \cite{2020arXiv200600219H} identified H$\alpha$ emitting sources in 14 BRCs based on slitless optical spectroscopy, including BRC 18. 

If these candidate YSOs are all formed as a result of triggering, and their motions are affected by the cloud's Rocket Effect, then they should be moving away from the exciting star, $\lambda$ Ori. In this work, we present results of the projected motion of the candidate YSOs associated with BRC 18 in order to investigate the possible ongoing Rocket Effect.

We organize the paper in the following manner. We provide description of the observational and archival data used in the paper in section \ref{sec:sel_brcs}, preliminary results and relevant discussions are presented in section \ref{sec:dis}, respectively. Finally, we conclude the work with a summary of the results in section \ref{sec:con}.
\section{Observation, Archival Data, and Data Analysis}\label{sec:sel_brcs}

\subsection{R-band polarimetric data}

\subsubsection{Instrumental setup and observing conditions}

We acquired optical R-band ($\lambda_{eff}=0.760~\mu$m) polarimetric data of 17 candidate YSOs using ARIES Imaging POLarimeter \citep[AIMPOL;][]{2004BASI...32..159R} attached with 1.04 m Sampurnanand Telescope, ARIES, Nainital, India. For observations, the central $325\times325$ pixel$^{2}$ area of a $1024\times1024$ pixel$^{2}$ charge-coupled device (CCD) chip (Tektronix TK1024) was used. The plate scale of the CCD is 1.48$\arcsec$ pixel$^{-1}$, and the field of view is $\sim8\arcmin$. The read-out noise and the gain of this CCD are 7.0 e$^{-1}$ and 11.98 e$^{-1}$ per Analog to Digital Unit, respectively. We observed 11 of the brightest (V$\leq$15) sources from the 32 targets listed as BRC 18 members in \cite{2001AJ....121.2124D}. These were supplemented with 6 additional targets found recently by Neha et al. whilst mapping the magnetic field geometry of BRC 18. The observations were taken over 8 nights in October and November 2016, as detailed in Table \ref{tab:log_obs_brc18}, where the six November observations were on consecutive nights starting on November 22. Typical air-mass and seeing during observations were $\lesssim1.4$ and $\lesssim2^{\prime\prime}$, respectively.

\begin{table}
	\centering
	\caption{Log of R-band polarimetric observations in the year of 2016.}
	\label{tab:log_obs_brc18}
	\begin{tabular}{lc}
		\hline
		Month (Date) & Heliocentric Julian Date\\\hline
        October 26 & 2457688.2964\\
        October 28 & 2457690.2963\\
        November 22 & 2457715.2945\\
        November 23 & 2457716.2944\\
        November 24 & 2457717.2943\\
        November 25 & 2457718.2942\\ 
        November 26 & 2457719.2941\\
        November 27 & 2457720.2940\\
        \hline    
	\end{tabular}\\
\end{table}

\subsubsection{Measuring polarization parameters}

The observed stellar image profile has a full width at half maximum (FWHM) $\sim2-3$ pixels. Only linear polarization can be obtained from AIMPOL, which contains a half-wave plate (HWP, acting as a modulator) and a Wollaston prism (acting as a beam splitter). This setup produces pairs of ordinary ($I_{o}$) and extraordinary ($I_{e}$) images of each target star on the CCD frame. During observations, the HWP is rotated at four positions, i.e. 0$\degr$, 22.5$\degr$, 45$\degr$, and 67.5$\degr$ to get four normalized Stokes parameters, which are q[R(0$\degr$)], u[R(22.5$\degr$)], q$_{1}$[R(45$\degr$)] and u$_{1}$[R(67.5$\degr$)], respectively. The errors in the normalized Stokes parameters were estimated following the relations provided by \cite{1998A&AS..128..369R}. 

The ratio R($\alpha$) is expressed as following:
\begin{equation}\label{eq1}
	R\left(\alpha\right)=\frac{\frac{I_{e}\left( \alpha\right)}{I_{o}\left( \alpha\right)}-1}{\frac{I_{e}\left(\alpha\right)}{I_{o}\left(\alpha\right)}+1} = Pcos\left(2\theta - 4\alpha\right),
\end{equation}
\noindent
where P is the fraction of total linearly polarized light and $\theta$ is the polarization angle. $\alpha$ is the angle of the fast axis of the HWP at four positions, 0$\degr$, 22.5$\degr$, 45$\degr$, and 67.5$\degr$ corresponding to four normalized Stokes parameters, q[$R$(0$\degr$)], u[$R$(22.5$\degr$)], q$_{1}$ [$R$(45$\degr$)], and u$_{1}$ [$R$(67.5$\degr$)], respectively. P and $\theta$ can be estimated from $\sqrt{q^2 + u^2}$ and 0.5 tan$^{-1}(u/q)$, respectively. By convention, $\theta$ is 0$\degr$ towards the north celestial pole and increases towards the east. The uncertainties in normalized Stokes parameters $\Delta{R\left(\alpha\right)}$ ($\Delta{q}$, $\Delta{u}$, $\Delta{q_{1}}$, and $\Delta{u_{1}}$) were computed using the expression below:
\begin{equation}\label{eq2}
\Delta{R\left(\alpha\right)}=\frac{\sqrt{I_{e}+I_{o}+2I_{b}}}{I_{e}+I_{o}},
\end{equation}
\noindent
where $I_{b}$[=$\frac{I_{be}+I_{bo}}{2}$] is the average background count surrounding the $I_{e}$ and $I_{o}$ pairs of the observed sources. The uncertainties in P\% and $\theta$ are computed using the equations below:
\begin{equation}
\Delta{P} =\frac{1}{P}\times\sqrt{q^{2}\Delta{q}^{2}+u^{2}\Delta{u}^{2}}, \hspace{0.5cm}\Delta{\theta} = \frac{1}{2P^{2}}\times\sqrt{q^{2}\Delta{u}^{2}+u^{2}\Delta{q}^{2}}\hspace{0.2cm}rad.
\end{equation}
Though the Stokes parameters $Q$ and $U$ can show both positive and negative values, the polarization P estimated from these parameters is always positive. Therefore, P always has a positive bias, especially in case of the sources with low signal-to-noise ratio. In order to remove this bias, we computed the debiased P\% using P$=\sqrt{\mathrm{P}^{2}-\Delta{P}^{2}}$, where $\Delta{P}$ is the error in P\% \citep{1974ApJ...194..249W, 2006PASP..118.1340V}.

After bias subtraction and flat-field correction, following the formula given by \cite{1998A&AS..128..369R}, we aligned and combined the multiple observed frames. The pair of $I_{o}$ and $I_{e}$ images of the individual star was selected using a program written in the Python language. Photometry of the selected pairs was performed using the Image Reduction and Analysis Facility (IRAF) DAOPHOT package to estimate the degrees of polarization (P\%) and position angle ($\theta$) of each target. The details of AIMPOL and steps of the data reduction procedure are given in \cite{2011MNRAS.411.1418E, 2012MNRAS.419.2587E, 2016A&A...588A..45N, 2018MNRAS.476.4442N, 2013MNRAS.432.1502S, 2015A&A...573A..34S, 2017MNRAS.465..559S}. The instrumental polarization from the polarization measurements was eliminated by observing some unpolarized standard stars from \cite{1992AJ....104.1563S}. In order to estimate the reference direction of the HWP during the observing runs, we observed six polarized standard stars (HD 236633, HD 15445, BD+59$\degr$389, HD 204827, HD 19820, HD 25443) from the list provided by \cite{1992AJ....104.1563S}. The measurements were used to estimate the zero-point offset with respect to the north, which was used to correct the position angles of the targets.

\subsection{Archival Data} \label{sec:gaia}

\textit{Gaia} EDR3 \citep{2020arXiv201201533G} provides astrometric measurements of 1.8 billion objects with unprecedented precision, brighter than $G$-band magnitude of 21. We obtained distance (\textit{d}) and proper motions in right ascension ($\mu_{\alpha\star}$) and declination ($\mu_{\delta}$) from \cite{2021AJ....161..147B} and the \textit{Gaia} EDR3 database \citep{2020arXiv201201533G}, respectively, by making a search around each source within a search radius of 1$\arcsec$. The initial criteria for inclusion of the sources were that $d/\Delta d$, $\mu_{\alpha\star}/\Delta\mu_{\alpha\star}$, and $\mu_{\delta}/\Delta\mu_{\delta}$ all must be $\gtrsim3$, where $\Delta$ represents the measurement uncertainties of the respective parameters. Then we included only those sources for which renormalized unit weight error (RUWE) $\lesssim1.4$ \citep{LL:LL-124}, indicating their reliable astrometric measurements. The candidate YSOs selected in our analysis have typical precision in $d$, $\mu_{\alpha\star}$, and $\mu_{\delta}$ are 2.0\%, 2.4\%, and 1.4\%, respectively.

We obtained Pan-STARRS (Panoramic Survey Telescope and Rapid Response System) \textit{g} and \textit{WISE W}1 counterparts of the candidate YSOs from \cite{2016arXiv161205560C} and \cite{2010AJ....140.1868W}, by giving a search radius of 1$\arcsec$ and 3$\arcsec$, respectively. In order to find the IR excess of the young sources, we acquired their 2MASS (Two Micron All Sky Survey) counterparts from \cite{2006AJ....131.1163S}, by giving a search radius of 1$\arcsec$. 
\vspace{-0.47cm}
\section{Results and Discussions}\label{sec:dis}

\subsection{Polarization of the candidate YSOs}\label{subsec:res_brc18_pol}

The polarimetric observations presented in this study were carried out before the data from the \textit{Gaia} were available (Table \ref{tab:log_obs_brc18}). Here, the latest \textit{Gaia} EDR3 measurements have been used  subsequently to combine the astrometric properties of the candidate YSOs along with their R-band polarimetric results. The polarimetric measurements and \textit{Gaia} astrometric and photometric results of the sources are provided in Table \ref{tab:sources_p_pa_new}. The candidate YSOs listed in the upper section of Table \ref{tab:sources_p_pa_new} are having reliable counterparts in \textit{Gaia} EDR3. The observed sources listed in the lower section of Table \ref{tab:sources_p_pa_new} do not have any reliable counterpart in \textit{Gaia} EDR3. The RUWE of these sources are higher than 1.4, indicating that their astrometric measurements might be unreliable. Therefore, we separated out this group of observed candidate YSOs. The P\% values range from 0.4 to 4.1. However, in terms of estimated P\% and $\theta$ values, there is no noticeable difference between the sources having reliable and unreliable counterparts in \textit{Gaia} EDR3.

\subsubsection{Finding candidate YSOs as BRC 18 members with \textit{Gaia} EDR3}

\begin{landscape}
    \begin{table}
    \scriptsize
		\centering
		\caption{Results of R-band polarization measurements of the observed candidate YSOs towards BRC 18 along with the astrometric and photometric properties obtained from \textit{Gaia} EDR3.}
		\label{tab:sources_p_pa_new}
		\renewcommand{\arraystretch}{1.5}
		\begin{tabular}{ccccccccccccccc}			
			\hline
			Serial & Source ID & RA$^{*}$ & Dec$^{*}$ & $\Delta$RA$^{\dagger}$ & $\Delta$Dec$^{\dagger}$ & P$\pm\Delta$P & $\theta$$\pm$$\Delta\theta$ & $d$$\pm$$\Delta d$ & $\mu_{\alpha\star}$$\pm$$\Delta\mu_{\alpha\star}$ & $\mu_{\delta}$$\pm\Delta$$\mu_{\delta}$ & RUWE & $G$$\pm\Delta$$G$$^{\ddagger}$&$G_\mathrm{BP}$$\pm$$\Delta G_\mathrm{BP}$$^{\ddagger}$ & $G_\mathrm{RP}$$\pm$$\Delta G_\mathrm{RP}$$^{\ddagger}$\\
			No.& (\textit{Gaia} EDR3) & ($^{\circ}$) & ($^{\circ}$) & ($^{\prime\prime}$) & ($^{\prime\prime}$) & (\%) & ($^{\circ}$) & (pc) & (mas yr$^{-1}$) & (mas yr$^{-1}$)&&(mag)&(mag)&(mag)\\ 
			(1)&(2)&(3)&(4)&(5)&(6)&(7)&(8)&(9)&(10)&(11)&(12)&(13)&(14)&(15)\\
			\hline
			\multicolumn{15}{c}{Sources having reliable counterparts in \textit{Gaia} EDR3}\\
			1 &3336149519712868736 &85.837176 & 9.101955 & 7287.9 &-2995.9&0.4$\pm$0.2 & 177$\pm$9 & 402$_{-3}^{4}$ & 2.145$\pm$0.021 & -2.665$\pm$0.015 &1.376& 12.581$\pm$0.012&13.343$\pm$0.043 & 11.691$\pm$0.030\\
			2 &3336159071720837120 &86.008674 &9.260364 & 7895.0 &-2425.6&2.3$\pm$0.3& 95$\pm$3 & 387$_{-5}^{5}$& 2.257$\pm$0.031 & -2.482$\pm$0.022 &1.012& 15.074$\pm$0.004&16.394$\pm$0.014 & 13.931$\pm$0.008\\	
			3 &3336107772631258624 &86.030253 & 9.110574 & 7973.3 &-2964.9&0.6$\pm$0.2 & 176$\pm$7 &393$_{-3}^{4}$ &1.661$\pm$0.031 &-2.703$\pm$0.018 &1.381& 13.694$\pm$0.004&14.458$\pm$0.012& 12.794$\pm$0.008\\
			4 &3336109563633776896 &86.044381 & 9.212530 & 8022.3 &-2597.8&0.7$\pm$0.2 & 182$\pm$10 & 396$_{-3}^{3}$ & 2.595$\pm$0.019 & -2.488$\pm$0.012 &1.094& 13.798$\pm$0.003&14.499$\pm$0.004& 12.985$\pm$0.004\\
			5 &3336104581471997440 &86.080668 & 9.089784 & 8152.6 &-3039.7&0.7$\pm$0.4& 147$\pm$11& 383$_{-5}^{5}$& 2.702$\pm$0.044 & -2.640$\pm$0.027 &1.059& 15.461$\pm$0.003&16.818$\pm$0.006& 14.313$\pm$0.004\\
			6 &3336108361043206912 &86.091194 & 9.147915 & 8189.2 &-2830.4&1.6$\pm$0.3 & 67$\pm$6 & 393$_{-3}^{4}$ & 2.250$\pm$0.028 & -2.483$\pm$0.018 &1.122& 14.513$\pm$0.003&15.707$\pm$0.008& 13.389$\pm$0.008\\
			7 &3336105028148587904 &86.108237 & 9.116840 & 8250.1 &-2942.3&0.7$\pm$0.2 & 179$\pm$10 &397$_{-2}^{3}$ & 2.571$\pm$0.018 & -2.720$\pm$0.011 &1.057& 13.645$\pm$0.008&14.488$\pm$0.026& 12.710$\pm$0.023\\	
			8 &3336204155993987200 &86.220057 & 9.218111 & 8645.8 &-2577.7&2.4$\pm$0.2 & 83$\pm$4 & 391$_{-2}^{4}$ & 1.909$\pm$0.022 & -2.321$\pm$0.014 &1.153& 14.141$\pm$0.005&14.968$\pm$0.014& 13.230$\pm$0.013\\
			9 &3336093513339795200 &86.267100 & 9.081243 & 8814.6 &-3070.5&1.0$\pm$0.3 &128$\pm$7 &394$_{-5}^{6}$ &2.323$\pm$0.038 & -2.859$\pm$0.022 &0.960& 15.178$\pm$0.004&16.482$\pm$0.011& 14.040$\pm$0.009\\
			\hline
			\multicolumn{15}{c}{Sources having unreliable counterparts in \textit{Gaia} EDR3}\\
			10 &3336149455290092672 &85.846768 & 9.092230 & 7322.1 &-3030.9&0.4$\pm$0.2 & 140$\pm$6 & 398$_{-37}^{46}$ & 1.522$\pm$0.163 & -1.206$\pm$0.134 & 6.156 & 13.395$\pm$0.019 & 13.632$\pm$0.051 & 11.913$\pm$0.040\\
			11 &3336149489649827840 &85.853067 & 9.102319 & 7344.3 &-2994.6&0.4$\pm$0.2 & 155$\pm$9 & 413$_{-8}^{8}$ & 0.809$\pm$0.043 & -2.220$\pm$0.034 & 2.772 & 12.770$\pm$0.003 & 13.482$\pm$0.005 & 11.947$\pm$0.005\\		
			12 &3336109116957457792 &86.037486 & 9.154068 & 7998.5 &-2808.3&1.3$\pm$0.4 & 176$\pm$11 & 396$_{-3}^{3}$ & 1.814$\pm$0.028 & -2.168$\pm$0.016 & 1.574 & 13.299$\pm$0.023 & 14.117$\pm$0.082 & 12.422$\pm$0.058\\
			13 &3336109288756140416 &86.072310 & 9.183034 & 8121.8 &-2704.0&0.8$\pm$0.2 & 137$\pm$4 & 541$_{-22}^{17}$ & 2.018$\pm$0.080 & -3.841$\pm$0.050 & 3.196 & 13.621$\pm$0.007 & 14.320$\pm$0.023 & 12.466$\pm$0.013\\		
			14 &3336111002446797184 &86.080096 & 9.209678 & 8149.1 &-2608.1&0.7$\pm$0.2 & 175$\pm$10 & 374$_{-6}^{6}$ & 1.937$\pm$0.043 & -2.107$\pm$0.027 & 2.285 & 14.006$\pm$0.005 & 15.147$\pm$0.018 & 12.932$\pm$0.011\\
			15 &3336110727568247552 &86.096725 & 9.201062 & 8208.2 &-2639.1&4.1$\pm$0.1 & 18$\pm$6 & 392$_{-3}^{2}$ &  2.980$\pm$0.024 & -2.365$\pm$0.015 & 1.453 & 11.757$\pm$0.003 & 12.315$\pm$0.005 &11.034$\pm$0.005\\
			16 &3336253530937700736 &86.186075 & 9.388527 & 8523.1 &-1964.2&0.6$\pm$0.2 & 164$\pm$8 & 327$_{-23}^{25}$ & 0.222$\pm$0.140 & 1.252$\pm$0.130 & 3.682 & 14.426$\pm$0.013 & 14.899$\pm$0.035 & 12.969$\pm$0.022\\			
			17 &3336093826873858304 &86.339785 & 9.095995 & 9072.5 &-3017.4&1.3$\pm$0.4 & 111$\pm$6 & 368$_{-8}^{9}$ &  1.306$\pm$0.068 & -2.723$\pm$0.044 & 1.932 & 15.336$\pm$0.003 &  16.926$\pm$0.006 & 14.093$\pm$0.005\\
			\hline
		\end{tabular}\\
		{\footnotesize $^{*}$ RA and Dec of the sources are in J2016 epoch obtained from \textit{Gaia} EDR3. $^{\dagger}$ Angular distance of the sources in RA and Dec from $\lambda$ Ori. $^{\ddagger}$ $G$, $G_\mathrm{BP}$, and $G_\mathrm{RP}$ magnitudes obtained from \textit{Gaia} EDR3, presented upto 3 decimal places.}
\end{table}

\end{landscape}

The ionizing source, $\lambda$ Ori is believed to be responsible for the bright rim seen towards BRC 18 \citep{1991ApJS...77...59S}. We adopted a distance and an age of $\sim$400 pc and $\sim5$ Myr, respectively, for the $\lambda$ Ori cluster \citep{2018AJ....156...84K} and a typical isothermal sound speed in H{\sc ii} region of 11.4 km s$^{-1}$ \citep{1989ApJ...346..735B,1990ApJ...354..529B} to estimate the time to reach the bright rim of BRC 18, which is 1.5 Myr. If the candidate YSOs associated with BRC 18 are all formed as a result of triggering due to the RDI caused by $\lambda$ Ori, then they are expected to have an age of approximately $\sim 5-1.5=3.5$ Myr. This age is consistent with the average age of the candidate YSOs distributed in the direction of BRC 18 estimated by \citet{2018AJ....156...84K} within the uncertainty. Then, assuming a typical velocity dispersion of $\sim$1 km s$^{-1}$ for the YSOs \citep{2009ApJ...703..399L,2015ApJ...815....2W}, within 3.5 Myr the maximum extent to which the YSOs could move from their place of birth is approximately 0.5$\degr$, considering the same distance (400 pc). Therefore, all the sources located within a circular area of radius 0.5$\degr$ centered around the \textit{IRAS} source (IRAS 05417+0907) are considered as a part of BRC 18.

\begin{figure*}
	\centering
	\includegraphics[height=11cm, width=16cm]{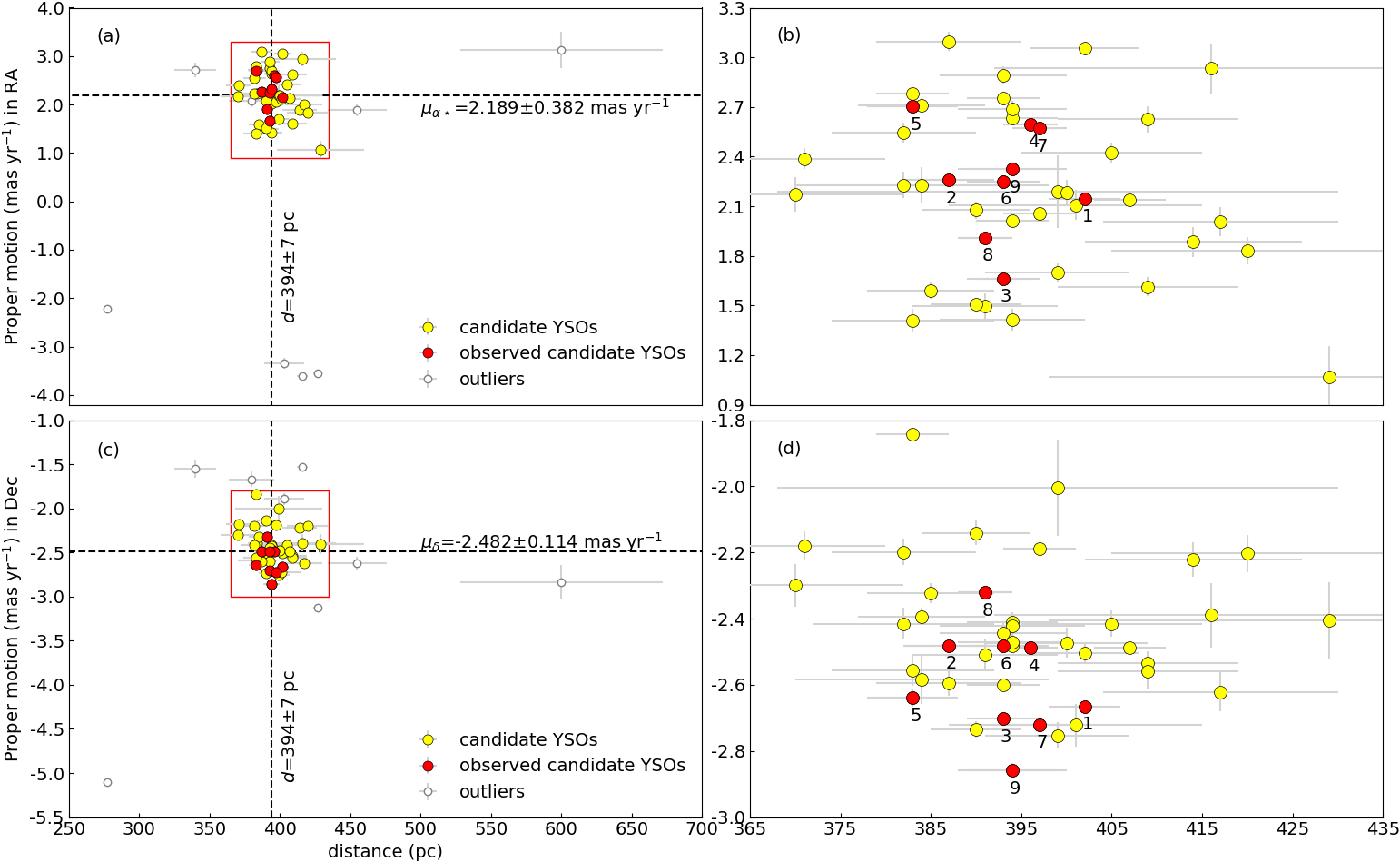}
	\caption{Proper motion values of the known candidate YSOs associated with BRC 18 are plotted as a function of their distances obtained from \textit{Gaia} EDR3. \textbf{(a)} $\mu_{\alpha\star}$ versus $d$ plot of the sources. The candidate YSOs lying within 5 $\times$ MAD boundary of $d$ versus $\mu_{\alpha\star}$ are shown using filled circles in yellow. The candidate YSOs having polarimetric observations and lying within the same 5 $\times$ MAD boundary are presented using filled circles in red. The outliers are shown using open circles. The dashed lines indicate the median values of $d$ and $\mu_{\alpha\star}$ of the candidate YSOs. \textbf{(b)} Zoomed in view of the box shown in (a), where only the sources lying within 5 $\times$ MAD are shown. The nine candidate YSOs having polarimetric measurements are labelled with their serial numbers (as listed in the upper section of Table \ref{tab:sources_p_pa_new}). \textbf{(c)} $\mu_{\delta}$ versus $d$ plot of the sources. The candidate YSOs lying within 5 $\times$ MAD boundary of $d$ versus $\mu_{\delta}$ are shown using filled circles in yellow. The candidate YSOs having polarimetric observations and lying within the same 5 $\times$ MAD boundary are presented using filled circles in red. The outliers are shown using open circles. The dashed lines indicate the median values of $d$ and $\mu_{\delta}$ of the candidate YSOs. \textbf{(d)} Zoomed in view of the box shown in (c), where only the sources lying within 5 $\times$ MAD are shown. The sources having polarimetric measurements are labelled here too.}
	\label{fig:br18_pm_dist}
\end{figure*}

\subsubsection{Kinematics of the candidate YSOs in BRC 18}

We found a total of 93 candidate YSOs within a circular region of radius 0.5$\degr$ centered around the \textit{IRAS} source in BRC 18 \citep{2001AJ....121.2124D, 2012PASJ...64...96H, 2015AJ....150..100K, 2018AJ....156...84K, 2018A&A...620A.172Z, 2020arXiv200600219H}. Of the 93, we obtained the \textit{Gaia} data for 53 sources after applying all the criteria listed in section \ref{sec:gaia}. In Fig. \ref{fig:br18_pm_dist} (a) and (c), we show the $\mu_{\alpha\star}$ and $\mu_{\delta}$ of the sources as a function of their distances ($d$), respectively, using filled circles in yellow. The enlarged views of Fig. \ref{fig:br18_pm_dist} (a) and (c) are presented in (b) and (d), respectively. The variance and the standard deviation that are commonly used in measuring the spread in a data set are more affected by the extreme (high and low) values. Hence, here we used median absolute deviation (MAD) to estimate the statistical dispersion in our data sets. The median values of $d$, $\mu_{\alpha\star}$, and $\mu_{\delta}$ are 394 pc, 2.170, and -2.482 mas yr$^{-1}$, respectively with corresponding MAD values as 10 pc, 0.455, and 0.159 mas yr$^{-1}$, respectively. We selected the sources lying within 5 $\times$ MAD with respect to the median values of the distance and the proper motions as sources that are associated with BRC 18. Choice of 5 $\times$ MAD was made based on the fact that the majority of the sources are lying within this extent. Also, the 5 $\times$ MAD boundary is roughly equivalent to a 3$\sigma$ limit, since 1$\sigma$ = 1.48 $\times$ MAD. A total of 43 sources satisfy this condition, and the remaining 10 are considered as outliers (shown in Fig. \ref{fig:br18_pm_dist} (a) and (c) using open circles). Table \ref{tab:yso_gaia} lists the astrometric properties of the 34 candidate YSOs, while the nine polarimetrically observed sources having reliable \textit{Gaia} EDR3 counterparts are presented in Table \ref{tab:sources_p_pa_new}. We recomputed the median and the MAD values for the 43 sources after eliminating the outliers to get the final astrometric results of the candidate YSOs in BRC 18. The new median values for $d$, $\mu_{\alpha\star}$, and $\mu_{\delta}$ are 394 pc, 2.189, and -2.482 mas yr$^{-1}$, respectively. The MAD values of the same are estimated as 7 pc, 0.382, and 0.114 mas yr$^{-1}$, respectively.

\subsubsection{Details of polarization results of candidate YSOs in BRC 18}

\begin{figure}
	\centering
	\includegraphics[height=13.5cm, width=8.5cm]{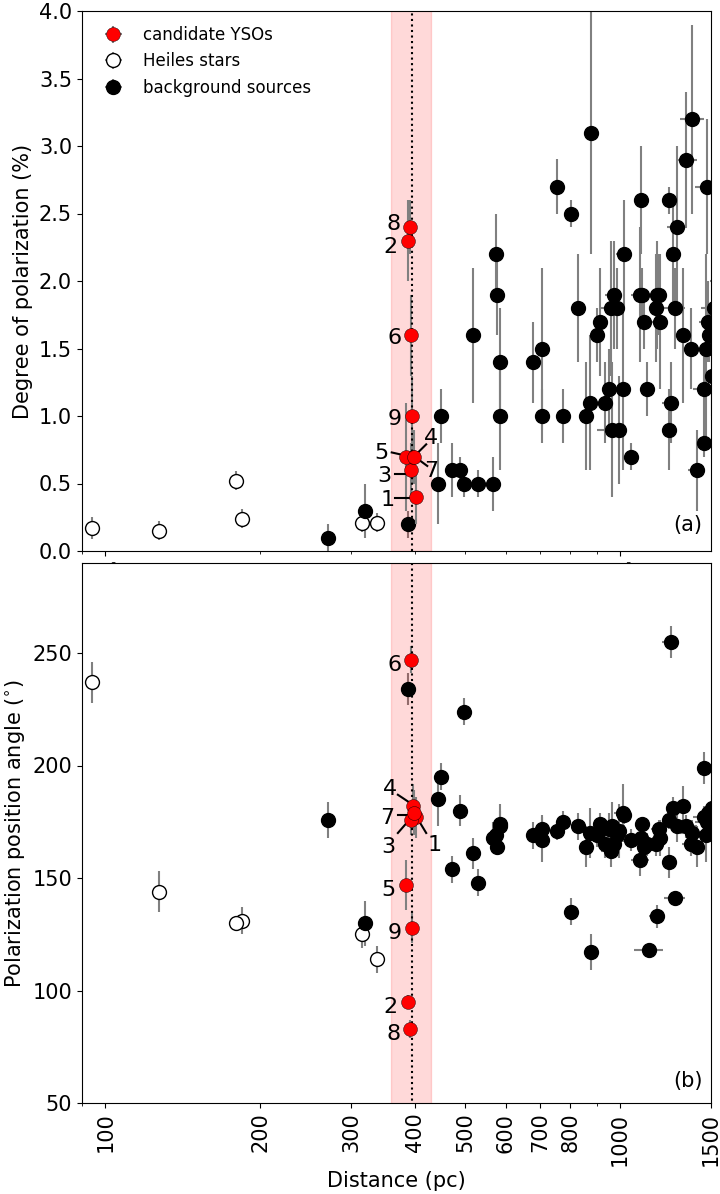}
	\caption{\textbf{(a)} The foreground corrected P\% versus distance of the sources (observed by Neha et al.) projected towards BRC 18 are shown using filled circles in black. The observed candidate YSOs are shown using filled circles in red. The six foreground sources are presented using open circles in black. The distance of 394 pc (distance of BRC 18) is shown using a dotted vertical line. The pink patch represents the cloud extent, which is 5$\times$MAD from 394 pc. \textbf{(b)} Polarization position angle versus distance plot for the same sources. Symbols represent the same as in (a).}
	\label{fig:yso_p_theta_dist}
\end{figure}

We made R-band polarization measurements of 17 candidate YSOs. Out of the 17, nine sources, which have reliable detections in \textit{Gaia} EDR3, are identified using filled circles in red, in Fig. \ref{fig:br18_pm_dist} (a) to (d). The measured polarization values have contributions from the dust grains located within the pencil beam along the line of sight of the stars, which could be the foreground, located in the cloud, and in the circumstellar material. As the cloud is located at $\sim$400 pc away from the Sun, it is important to remove the contribution of the foreground material from the measured polarization values. For this, six foreground stars (HD 38527, HD 39007, HD 37408, HD 37355, HD 38096, and HD 37522), located within a circle of radius 2$\degr$ centered around IRAS 05417+0907 were selected. The distances of these sources range from $\sim 94-334$ pc \citep{2021AJ....161..147B}. In Fig. \ref{fig:yso_p_theta_dist}, the P\% versus distance and the $\theta$ versus distance plots for the six foreground sources are shown using open circles. The mean values of the P\% and $\theta$ for these six stars are 0.2\% and 117$\degr$, respectively, with corresponding standard deviations as 0.1\% and 28$\degr$, respectively. The foreground contribution to the measured polarization was subtracted vectorially. No significant change was noticed as expected since the foreground polarization is found to be very low. 

\begin{figure}
	\centering
	\includegraphics[height=7.5cm, width=8.5cm]{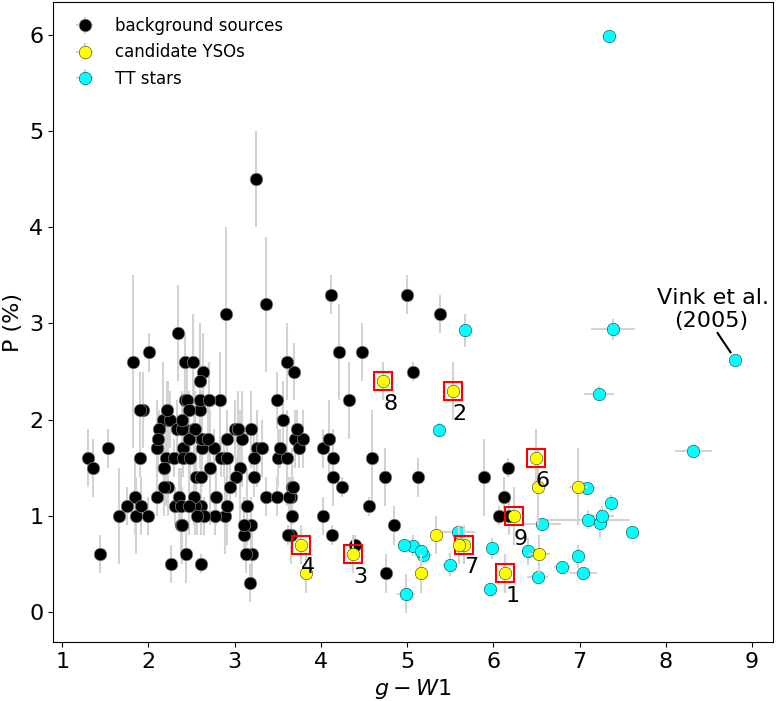}
	\caption{Plot of the P\% versus Pan-STARRS $g$ - \textit{WISE W}1 colors of the sources. The filled circles in black indicate the sources projected background of BRC 18 (observed by Neha et al.), while the filled circles in yellow are the candidate YSOs observed by us in polarimetric mode. The candidate YSOs having reliable \textit{Gaia} EDR3 data are marked with open square boxes in red. The filled circles in cyan are the 28 TT stars (27 are obtained from \citet{1982A&AS...48..153B} and one is taken from \citet{2005MNRAS.359.1049V}, which is labelled here.}
	\label{fig:br18_pol_gw1}
\end{figure}

In Fig. \ref{fig:yso_p_theta_dist} (a) and (b), we show the foreground corrected P\% and $\theta$ as a function of distances of the 64 sources, which are obtained from \cite{2021AJ....161..147B}, using filled circles in black. We obtained distance measurements of 146 sources, which were observed by Neha et al. in polarimetric mode and in Fig. \ref{fig:yso_p_theta_dist} (a) and (b), we only show the sources lying within 1.5 Kpc. Because the majority of these sources are lying behind the cloud, the measured $\theta$, most likely, is representing the orientation of the cloud's magnetic field. Two sources lying at 270 pc and 320 pc show very low P\% values, consistent with the values obtained for the six foreground sources. Filled circles in red represent polarimetric results of the nine candidate YSOs having reliable detections in \textit{Gaia} EDR3. The mean values of the P\% and $\theta$ for these nine sources are 1.2\% and 157$\degr$, respectively, with corresponding standard deviations as 0.7\% and 47$\degr$, respectively. The dotted line represents the median distance of 394 pc, and the shaded portion in pink represents five times of the MAD value. 

Unpolarized starlight, when travels through a molecular cloud, gets polarized due to the selective absorption by the dust grains present in that cloud. Normally, the gradual increase in P\% is expected as a function of distance due to the increasing column of dust grains along the path of the starlight. But when the path of the starlight passes through any molecular cloud, a sudden rise in the values of P\% occurs due to the additional polarization caused by the dust grains residing in the cloud. The stars that are background to the molecular cloud are expected to show relatively higher values of P\%, compared to those lying foreground to it. The distance at which P\% rises significantly, is considered as the distance of the cloud \citep[e.g.][]{1992BaltA...1..149S, 1997A&A...327.1194W, 1998A&A...338..897K, 2007A&A...470..597A}. The sudden rise in P\% of both the candidate YSOs and the normal stars in Fig. \ref{fig:yso_p_theta_dist} (a) also implies that BRC 18 is located at a distance of $\sim394$ pc. Only one observed source (not identified as candidate YSO) located at the pink patch is having $\mu_{\alpha\star}$ and $\mu_{\delta}$ as -17.048$\pm$0.019 and 2.414$\pm$0.011 mas yr$^{-1}$, respectively, indicating this source to be a field star.	

\subsubsection{Polarization and IR excess of the sources towards BRC 18}

In Fig. \ref{fig:br18_pol_gw1}, we show the P\% as a function of the ($g-W1$) colors. The $g$-magnitudes for the sources are obtained from the Pan-STARRS \citep[$\lambda_{eff}\sim4810$ \AA; ][]{2016arXiv161205560C} and the $W1$ values are obtained from the \textit{WISE} \citep[$\lambda_{eff}\sim3.6~ \mu$m; ][]{2010AJ....140.1868W}. The filled circles in black indicate the sources observed by Neha et al., which are likely to be the normal field stars, while the filled circles in yellow are the 15 of the 17 candidate YSOs, for which we have polarization measurements and $g-W1$ colors. The candidate YSOs, which have reliable \textit{Gaia} EDR3 data are distinguished by using open square boxes in red. In a study conducted to investigate the correlations between polarization and other observable properties, \citet{1982A&AS...48..153B} found correlations between polarization and average IR color indices, especially $V-L$. The observed correlation was attributed to the absorption of stellar radiation by the dust and re-emission in IR wavelengths. We took the polarization measurements of a number of T Tauri (TT) stars from \citet{1982A&AS...48..153B} made at 7543 \AA~ and plotted as a function of their $g-W1$ colors. Of the 55 TT stars studied by him, for 27, we obtained $g-W1$ colors, which are shown using filled circles in cyan in Fig. \ref{fig:br18_pol_gw1}. We also added polarimetric measurement of one TT star (HP Tau) from \cite{2005MNRAS.359.1049V}, observed in R-band (6500 \AA). It is also shown in Fig. \ref{fig:br18_pol_gw1} using same symbol (filled circle in cyan) and labelled separately. Majority of the 15 candidate YSOs observed by us are also falling in the similar region occupied by these TT stars. It is to be noted that the polarization and color measurements are not simultaneous, and young sources are known for their variability in both the P\% and magnitudes \citep[e.g., ][]{1987IAUS..122..135J, 2009A&A...499..137M, 2011ApJ...732...69L}. In spite of these caveats, we find that the TT stars and the 15 sources observed by us show higher $g-W1$ colors indicative of the presence of circumstellar material. The circumstellar dust distributed in a disc geometry can cause relatively large P\% values, which depends on the amount of dust scattering due to the radiation from the central star, degree of the disc flattening, and its orientation with respect to our line of sight to the star \citep[e.g., ][]{2016MNRAS.456.2794Y}. Therefore the variation of P\% seen in candidate YSOs could be possibly due to any of these factors or a combination of all.

\begin{figure}
	\centering
	\includegraphics[height=7.3cm, width=8.5cm]{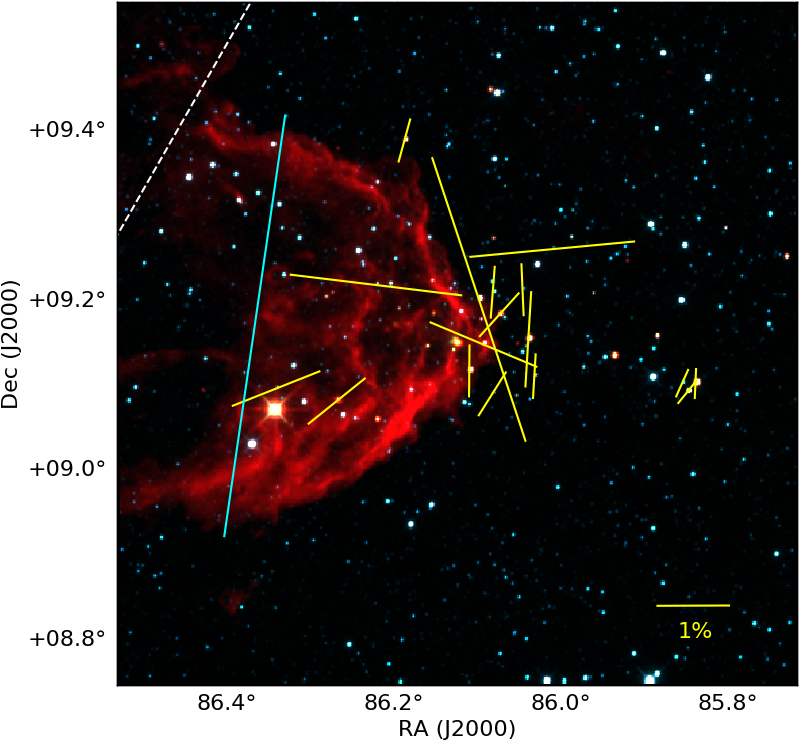} 
	\caption{Polarimetric results of the candidate YSOs are overplotted on the \textit{WISE} color composite diagram using 3.6 (blue), 4.5 (green), and 12 (red) $\mu$m images. The lines in yellow represent the polarization vectors of the observed candidate YSOs. The mean direction of the projected magnetic field (obtained by Neha et al.) is indicated by a line drawn in cyan. The white dashed line represents the Galactic plane. A vector with 1\% polarization is shown as a reference at the bottom right corner.}
	\label{fig:yso_p_theta_vector}
\end{figure}

\begin{figure}
	\centering
	\includegraphics[height=6cm, width=8.5cm]{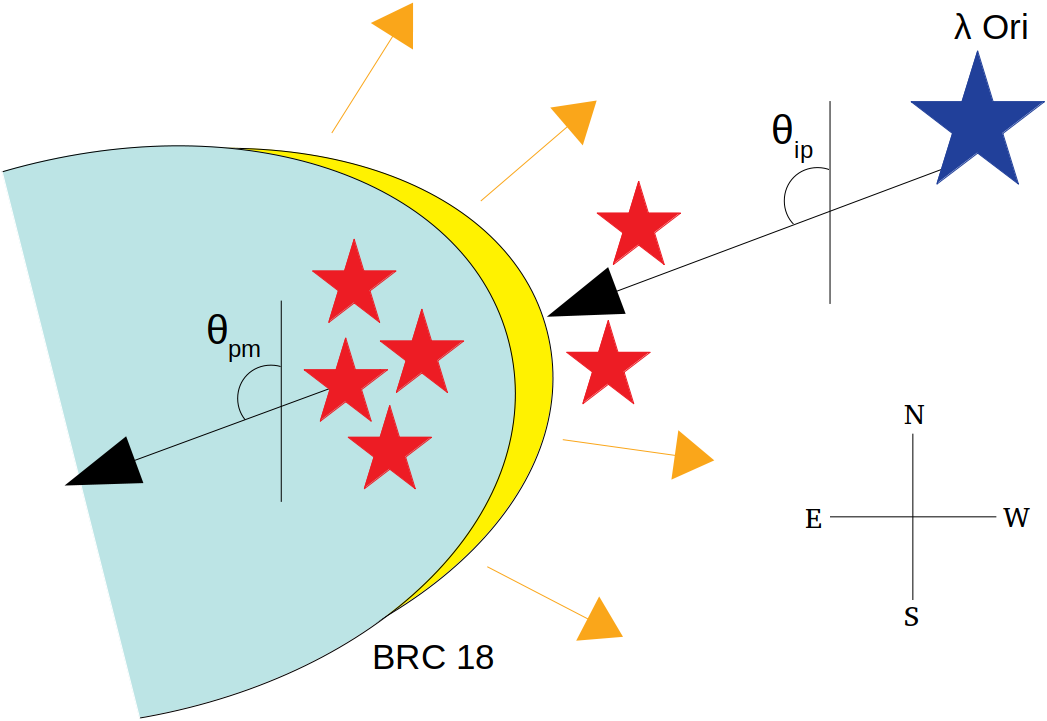}
	\caption{Cartoon diagram of a system consisting of BRC 18 and the main ionizing source $\lambda$ Ori (not to scale). The massive ionizing source is indicated by a blue star symbol, while the candidate YSOs projected towards BRC 18 are shown using red star symbols. The thick yellow border indicates the bright rim of the cloud. The direction of ionizing photons is shown using an arrow in black pointing away from the massive star, making the angle of ionizing photons $\theta_{ip}$ with respect to the north. The angle of the relative proper motion of candidate YSOs with respect to $\lambda$ Ori is presented by $\theta_{pm}$, which is made by another arrow in black in BRC 18. The arrows in yellow indicate the direction of photo-evaporating cloud material, as a reaction of which BRC 18 accelerates away from the ionizing source.}
	\label{fig:model_rdi}
\end{figure}

\subsubsection{Effect of magnetic field on the polarization of candidate YSOs}

As a group, the 17 candidate YSOs show a relatively higher degree of polarization with a mean value of 1.2\% when compared to the normal sources (observed by Neha et al.) lying at similar distances. The mean value of $\theta$ for the normal background sources is 172$\degr$, and the corresponding dispersion is 7$\degr$, which is small compared to the dispersion in $\theta$ estimated for the candidate YSOs. This suggests that the cloud has a well ordered magnetic field morphology, and the candidate YSOs are showing much diverse values of position angles. The projected magnetic field of BRC 18 is oriented roughly in the north-south direction at 172$\degr$ (starting from 0$\degr$ towards the north and increasing towards east), and the $\theta$ values for the candidate YSOs are relatively more random. Thus, there is no preferred orientation of their circumstellar discs (either parallel or perpendicular) with respect to the cloud's magnetic field. In Fig. \ref{fig:yso_p_theta_vector}, we present the polarization vectors of the candidate YSOs using vectors drawn in yellow and overplotted on the \textit{WISE} color composite diagram using 3.6 (blue), 4.5 (green), and 12 (red) $\mu$m images. The line drawn in cyan represents the mean position angle, i.e., the angle of projected magnetic field obtained by Neha et al. using the background sources. From Fig. \ref{fig:yso_p_theta_vector}, it is clear that the polarimetric results of the candidate YSOs are scattered with respect to the projected magnetic field direction. Our results are in agreement with the results obtained by \cite{2004A&A...425..973M} in Taurus, where they found that the $\theta$ values for the YSOs are irregular. One possibility for random orientations could be that the YSOs are formed with discs oriented in an arbitrary manner \citep{2004A&A...425..973M}. Additionally, they also mentioned that the irregularity could be originated from the projection effect. The third reason could be that the magnetic field strongly drives the gravitational collapse, turning all systems to be oriented parallel to the field lines, and some of these systems become misaligned later by some unidentified processes, for example, dynamical encounters \citep{2004A&A...425..973M, 2010MNRAS.401.1505B, 2016ApJ...828...48V}. 

\begin{figure*}
	\centering
	\includegraphics[height=6cm, width=\textwidth]{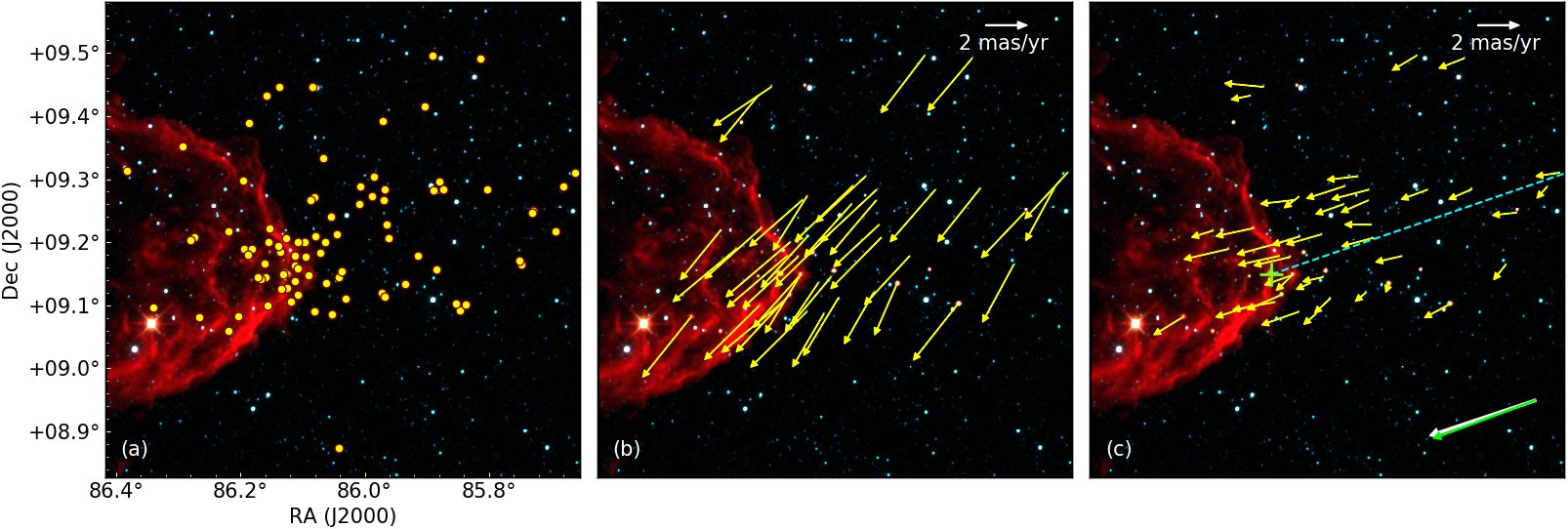}
	\caption{Proper motion properties of the candidate YSOs in BRC 18 on the \textit{WISE} color composite diagram using 3.6 (blue), 4.5 (green), and 12 (red) $\mu$m images. \textbf{(a)} The spatial distribution of the 93 known candidate YSOs associated with BRC 18 is shown using filled circles in yellow. \textbf{(b)} The arrows in yellow represent the proper motion vectors of the candidate YSOs obtained from \textit{Gaia} EDR3. \textbf{(c)} The arrows in yellow represent the relative proper motion vectors of the candidate YSOs with respect to the ionizing source. The dashed line in cyan represents the direction of the ionizing radiation with respect to the \textit{IRAS} source (`+' symbol in green) embedded in BRC 18. The median value of the relative proper motions of the candidate YSOs and the direction of ionizing photons are indicated using vectors in green and white, respectively, in the lower right corner (not to scale).}
	\label{fig:brc18_pm}
\end{figure*}

In the case of BRC 18, the energetic ionizing photons from $\lambda$ Ori could be expected to be responsible for the random orientations of the circumstellar discs of candidate YSOs. But then, the cloud's magnetic field lines also should be affected by the same. \cite{2009MNRAS.398..157H} presented the first three-dimensional magnetohydrodynamical (MHD) simulations of magnetized globules. They showed that these globules would evolve into flattened sheet-like structures if they have a strong magnetic field aligned perpendicular to the direction of ionization. \cite{2011MNRAS.412.2079M} showed that initially perpendicular weak and medium magnetic field lines are eventually aligned with the elongated structures of the evolved clouds. Based on the R-band polarimetric study of the stars located at the periphery of the entire cloud by Neha et al., the magnetic field strength towards BRC 18 is estimated as $\sim80~\mu$G, which is of medium strength, oriented almost perpendicular to the direction of ionization. The moderate strength of the magnetic field does not significantly influence the interaction of the cloud and ionization front, and the ionized photo-evaporation flow can escape out \citep{2011MNRAS.414.1747A}. Also, the effect of the magnetic field is not significant at small scales, e.g., stellar sizes, which could be responsible for the randomization of the circumstellar discs of the candidate YSOs.

\subsection{Projected motion of the candidate YSOs}\label{subsec:rel_kin_yso}

Proper motion of an H{\sc ii} region in a well-defined reference frame can be used to estimate its motion as a whole in the plane-of-sky. However, the proper motion values of individual stars subtracted from the mean proper motion of the H{\sc ii} region determines their internal projected motion \citep{1997MmSAI..68..833J}. We can consider the proper motion of the ionizing source as the proxy of the proper motion of its H{\sc ii} region. In Fig. \ref{fig:model_rdi}, we have shown a cartoon diagram of a system containing BRC 18 and the ionizing source $\lambda$ Ori. The massive ionizing source is shown using a star symbol in blue, while the YSOs projected towards the BRC are indicated by star symbols in red. The arrows in yellow indicate the direction of photo-evaporation flow of the cloud material. $\theta_{ip}$ and $\theta_{pm}$ represent the angle of ionizing photons and the angle of the internal projected motion of YSOs with respect to the north, respectively. Both $\theta_{ip}$ and $\theta_{pm}$ are measured from north and increasing eastward (anti-clockwise) as shown in Fig. \ref{fig:model_rdi}. 

In Fig. \ref{fig:brc18_pm} (a), we show the spatial distribution of the known candidate YSOs associated with BRC 18 using filled circles in yellow overplotted on \textit{WISE} color composite diagram using 3.6 (blue), 4.5 (green), and 12 (red) $\mu$m images. In Fig. \ref{fig:brc18_pm} (b), the arrows in yellow represent the observed proper motion vectors of the candidate YSOs obtained from \textit{Gaia} EDR3. In order to obtain the internal projected motion of the candidate YSOs in BRC 18, which is located at the eastern edge of the $\lambda$ Ori H{\sc ii} region, we first note the astrometric parameters of the earliest type ionizing source $\lambda$ Ori, considered to be mainly responsible for the ionization of the $\lambda$ Ori H{\sc ii} region. The \textit{Gaia} EDR3 provides $d$=$439_{-77}^{112}$ pc, $\mu_{\alpha\star}$ = 2.896$\pm$0.614 mas yr$^{-1}$, and $\mu_{\delta}$ = -3.183$\pm$0.425 mas yr$^{-1}$ of this star. However, RUWE for $\lambda$ Ori is relatively high (4.823). There could be a presence of unresolved binaries, which is responsible for such deviations \citep{2020arXiv201201533G}. Therefore it is possible that the present \textit{Gaia} EDR3 parallax measurement of $\lambda$ Ori may be uncertain. We, therefore, consider the median values of proper motions and distances of the spectroscopically confirmed members of the Collinder 69 cluster \citep{2012A&A...547A..80B}, of which $\lambda$ Ori is a part. The median $d$ is 392$\pm$8 pc, $\mu_{\alpha\star}$ = 0.919$\pm$0.204 mas yr$^{-1}$, and $\mu_{\delta}$ = -2.017$\pm$0.121 mas yr$^{-1}$, after removing the outliers. We have used these values as a reference in our analysis instead of the same for $\lambda$ Ori.

The median values of proper motion obtained above were subtracted from the observed proper motion values of the candidate YSOs to obtain their true internal projected motion or the relative proper motion with respect to the ionizing source. In Fig. \ref{fig:brc18_pm} (c), the arrows in yellow represent the relative proper motion vectors of the candidate YSOs with respect to the ionizing source $\lambda$ Ori. The dashed line in cyan represents the direction of the ionizing radiation with respect to IRAS 05417+0907 (`+' symbol in green) embedded in BRC 18. The median direction of the relative proper motions of the candidate YSOs and the direction of ionizing photons are indicated using vectors in green and white, respectively, in the lower right corner. Being formed inside the cloud, the YSOs share similar kinematics as the former. We consider the median angle made by the relative proper motion of the YSOs with respect to $\lambda$ Ori as $\theta_{pm}$, starting from the celestial N-S axis. The $\theta_{ip}$ has been computed as the angle of the line joining the ionizing source and the central \textit{IRAS} source located in BRC 18, with respect to the celestial N-S axis. Therefore, in an ideal Rocket Effect scenario in the sky plane, the absolute difference between $\theta_{ip}$ and $\theta_{pm}$ ($\mid\theta_{ip}-\theta_{pm}\mid$) would be $\simeq0\degr$ (see Fig. \ref{fig:model_rdi}). In BRC 18, we found $\mid\theta_{ip}-\theta_{pm}\mid$=$\mid109\degr-110\degr\mid$=$1\degr$ with an uncertainty of $14\degr$. 

Now, this is to note that in our analysis, we considered only the motions of the candidate YSOs projected in the sky plane, which is the proper motion. We did not consider the radial velocities of the candidate YSOs, which indicate the motion of these sources along the line of sight. If the massive source is located at relatively far behind or in front of the BRC (i.e., the massive source and the BRC are not lying on the sky plane but making a significant angle with respect to the sky plane), then the candidate YSOs associated with the BRC can have a significant radial component that would possibly be towards or away from us, respectively, depending on the relative distance between the ionizing source and the BRC. Therefore, the $\mid\theta_{ip}-\theta_{pm}\mid\sim$0$\degr$ condition does not completely indicate the Rocket Effect in 3D but in 2D. 

\begin{figure*}
	\centering
	\includegraphics[height=10.5cm, width=\textwidth]{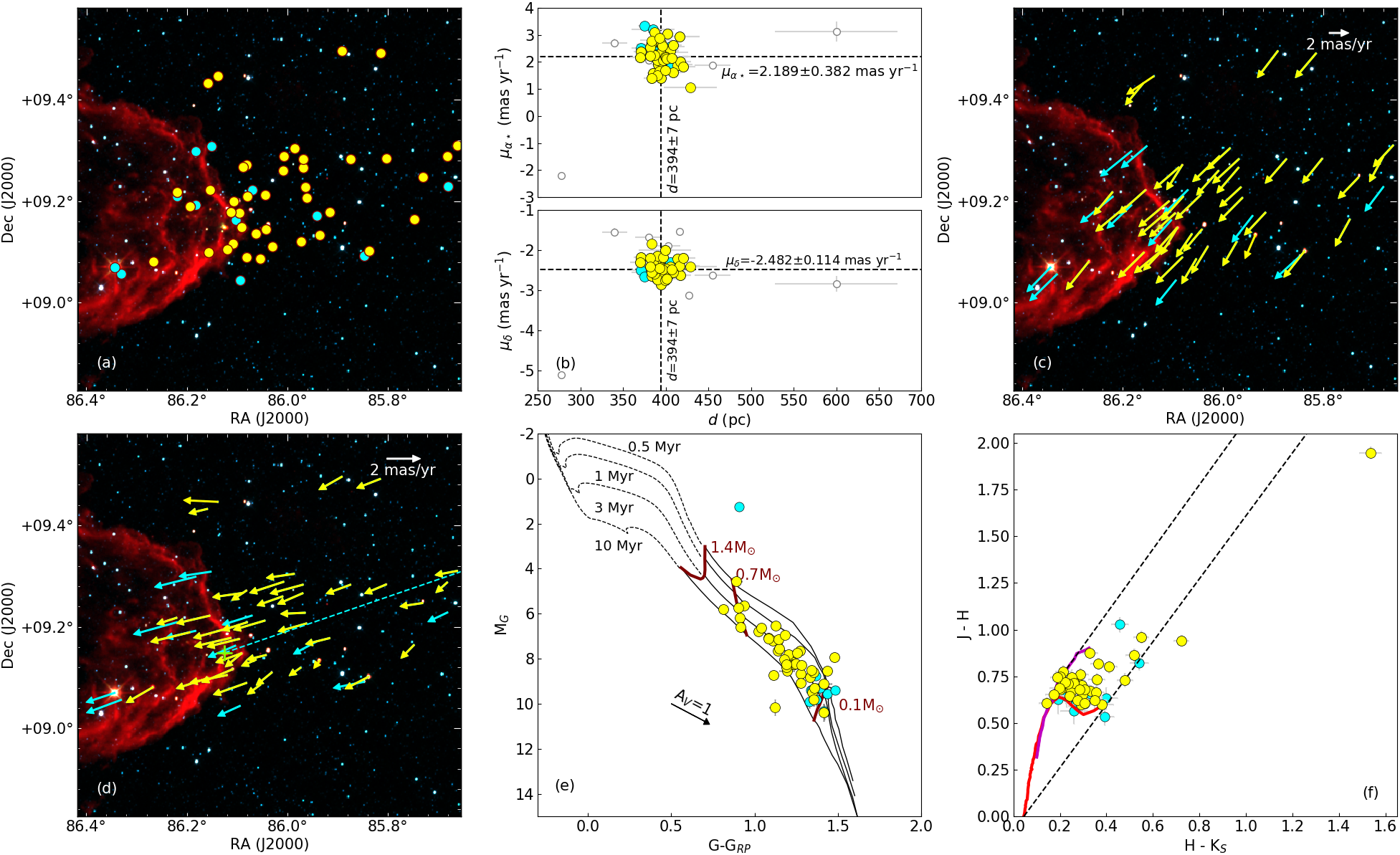}
	\caption{\textbf{(a)} The known candidate YSOs (filled circles in yellow) and comoving sources (filled circles in cyan) associated with BRC 18 overplotted on the \textit{WISE} color composite diagram using 3.6 (blue), 4.5 (green), and 12 (red) $\mu$m images. \textbf{(b) upper panel:} $\mu_{\alpha\star}$ of the known candidate YSOs and comoving sources are plotted as a function of their $d$ obtained from \textit{Gaia} EDR3, in filled circles in yellow and cyan, respectively. The open circles represent the $d$-$\mu_{\alpha\star}$ values of the outliers. The dashed lines show the median values of $d$ and $\mu_{\alpha\star}$ of the candidate YSOs. \textbf{(b) lower panel:} Same as (b) upper panel, but for $\mu_{\delta}$ versus $d$ of the sources. \textbf{(c)} The observed proper motion vectors of the known candidate YSOs (vectors in yellow) and the comoving sources (vectors in cyan) obtained from \textit{Gaia} EDR3. \textbf{(d)} The relative proper motion vectors of the known candidate YSOs (vectors in yellow) and the comoving sources (vectors in cyan) with respect to the central ionizing source. The dashed line in cyan represents the direction of the ionizing radiation with respect to IRAS 05417+0907 (shown using `+' symbol in green). \textbf{(e)} \textit{Gaia} M$_{G}$ versus ($G$-$G_{\mathrm{RP}}$) CMD of the candidate YSOs (filled circles in yellow) and comoving sources (filled circles in cyan). The dashed lines indicate the PMS isochrones from PARSEC models, and solid curves represent the same from CIFIST models. The PMS isochrones corresponding to 0.5, 1, 3, and 10 Myr are shown. A reddening vector corresponding to A$_{V}$=1 magnitude is also shown. \textbf{(f)} 2MASS $J-H$ versus $H-K_{s}$ color-color diagram of the candidate YSOs (filled circles in yellow) and comoving sources (filled circles in cyan). The solid curves in red and magenta represent the loci of the unreddened main sequence stars and giants, respectively. The dashed lines in black indicate the extinction vectors.}
	\label{fig:brc18_pm_comv}
\end{figure*}

\subsection{Search for additional comoving sources}\label{subsec:com_yso}

Based on the work done by \cite{2020MNRAS.494.5851S, 2021arXiv210412061S}, we looked for additional sources, which share kinematics similar to the previously known candidate YSOs in BRC 18. These sources could possibly be young stars not identified in earlier studies. We have found 12 such additional sources, i.e. the comoving sources. Table \ref{tab:comoving_gaia} lists the astrometric properties of these sources. All the comoving sources have m/$\Delta$m$\geq$3 and RUWE$\leq$1.4 and are lying within 5 $\times$ MAD boundaries with respect to the median values of $d$, $\mu_{\alpha\star}$, and $\mu_{\delta}$ of the known candidate YSOs. It is to be noted that the comoving sources identified are based on their distances and proper motions alone. Thus it is possible that at least a few of them could be field stars instead of being young sources. A detailed spectroscopic study of these sources could help to identify their true nature.

In Fig. \ref{fig:brc18_pm_comv} (a), we present the spatial distribution of candidate YSOs and comoving sources in BRC 18 using filled circles in yellow and cyan, respectively. Fig. \ref{fig:brc18_pm_comv} (b) shows $\mu_{\alpha\star}$ and $\mu_{\delta}$ of the known candidate YSOs (yellow) and comoving sources (cyan) as a function of their $d$ obtained from \textit{Gaia} EDR3. The $d$-$\mu_{\alpha\star}$ and $d$-$\mu_{\delta}$ values are shown in the upper and lower panels, respectively. The open circles represent the outliers. The dashed lines show the median values of $d$, $\mu_{\alpha\star}$, and $\mu_{\delta}$ of the candidate YSOs. The observed proper motion vectors of the known candidate YSOs (vectors in yellow) and the comoving sources (vectors in cyan) obtained from \textit{Gaia} EDR3 are shown in Fig. \ref{fig:brc18_pm_comv} (c) and Fig. \ref{fig:brc18_pm_comv} (d) presents the relative proper motion vectors of the known candidate YSOs (vectors in yellow) and the comoving sources (vectors in cyan) with respect to the central ionizing source $\lambda$ Ori. 

In Figure 7 (e), we present an absolute \textit{Gaia G} ($M_{G}$) versus $G-G_{RP}$ colour-magnitude diagram (CMD, not corrected for extinction/reddening) of the newly found comoving sources, overplotted with the candidate YSOs, using filled circles in cyan and yellow, respectively. From this figure, it is clear that the comoving sources have similar ages as the candidate YSOs. The PMS isochrones corresponding to 0.5, 1, 3, and 10 Myr are also shown. We used the CIFIST \citep[Cosmological Impact of the First STars; ][]{2015A&A...577A..42B} models for low-mass stars (thick curves in black) and the PARSEC \citep[PAdova and TRieste Stellar Evolution Code; ][]{2012MNRAS.427..127B} models for high-mass stars (dashed curves in black). A reddening vector corresponding to A$_{V}$=1 magnitude is also shown. 

We show 2MASS $J-H$ versus $H-K_{s}$ color-color diagram of candidate YSOs and comoving sources in Fig. \ref{fig:brc18_pm_comv} (f), using filled circles in yellow and cyan, respectively. The solid curves in red and magenta represent the loci of the unreddened main sequence stars and giants, respectively. The dashed lines in black indicate the extinction vectors. This plot clearly shows that the latter suffer from less extinction and less near-IR excess, although a fraction of them are found to have similar excess as the known candidate YSOs. This further confirms that the comoving sources are potential YSOs evolving along with the known candidate YSOs not identified in earlier studies. However, to confirm their youth, more detailed spectroscopic studies are being carried out currently. The results will be presented shortly in a subsequent paper.
 
\section{Summary and Conclusions}\label{sec:con}

We present R-band polarimetric results of 17 candidate YSOs projected towards BRC 18 to study their circumstellar disc orientations. Star formation in BRC 18 is considered to be affected by the impact of the ionizing source $\lambda$ Ori, which is called RDI. As a consequence of the RDI effect, the collapse of the BRCs leads to triggered star formation, leaving behind a chain of YSOs, as reminiscent of the elongated cloud. Furthermore, the cloud accelerates away from the direction of the ionizing source due to the photo-evaporation of cloud material towards the ionized medium, which is known as the ``Rocket Effect''. The fact that the internal projected motion of the YSOs mirrors the motion of the parental cloud in the plane-of-sky is investigated herein BRC 18 using the astrometric information of the candidate YSOs from \textit{Gaia} EDR3. The main conclusions of this work are summarized below:

\begin{itemize}
	\item Based on the \textit{Gaia} measurements of the candidate YSOs located towards BRC 18, we estimated the median distance to this cloud as 394$\pm$7 pc. The median $\mu_{\alpha\star}$ and $\mu_{\delta}$ of these sources are 2.189$\pm$0.382 and -2.482$\pm$0.114 mas yr$^{-1}$, respectively.
	\item The sudden rise in the polarization versus distance plot of the sources projected background to BRC 18 also indicates a similar distance to the cloud.
	\item We obtained the disc orientations of 17 candidate YSOs in BRC 18 to be randomly oriented, suggesting that the projected magnetic field of BRC 18 has more impact at a large scale, and its impact decreases at a smaller scale (scale of individual candidate YSOs), which leads to the possibility of the effect of close dynamical interactions on the disk orientations. Another probability is that the candidate YSOs could have changed their disc orientations since birth. Because of projection effect also, this randomness can appear.
	\item We found $\mid\theta_{ip}-\theta_{pm}\mid=1\degr\pm14\degr$ in BRC 18, i.e. the directions of the ionization front and the median internal projected motion of the candidate YSOs are almost parallel.
	\item We searched for sources moving in a similar fashion as the other candidate YSOs do. We obtained 12 such additional comoving sources located towards BRC 18. Most of them show a distribution similar to the candidate YSOs in the \textit{Gaia} CMD. 	
	\item Based on 2MASS $J-H$ versus $H-K_{s}$ color-color diagram, we found less extinction and little or no near-IR excess in these newly identified comoving sources.
\end{itemize}

The evidence of ``Rocket Effect'' found in BRC 18, based on relative internal projected motions of the associated candidate YSOs are being examined in other BRCs also. The results will be presented in a subsequent paper.
\section*{Acknowledgements}

We sincerely thank the referee, Dr. Alexander Slater Binks, for his wise and insightful suggestions that improved our manuscript significantly. We thank all the supporting staff at ARIES, Nainital, who made the polarimetric observations possible. We acknowledge the support by the S. N. Bose National Centre for Basic Sciences under the Department of Science and Technology, Govt. of India. This work has made use of archival data from the following sources: (1) European Space Agency (ESA) mission \textit {Gaia} (\url{https://www.cosmos.esa.int/gaia}), processed by the \textit{Gaia} Data Processing and Analysis Consortium (DPAC, \url{https://www.cosmos.esa.int/web/gaia/dpac/consortium}). Funding for the DPAC has been provided by national institutions, in particular the institutions participating in the {\it Gaia} Multilateral Agreement (MLA); (2) The Pan-STARRS1 Surveys were made possible through contributions by the Institute for Astronomy, the University of Hawaii, the Pan-STARRS Project Office, the Max-Planck Society and its participating institutes, the Max Planck Institute for Astronomy, Heidelberg and the Max Planck Institute for Extraterrestrial Physics, Garching, The Johns Hopkins University, Durham University, the University of Edinburgh, the Queen's University Belfast, the Harvard-Smithsonian Center for Astrophysics, the Las Cumbres Observatory Global Telescope Network Incorporated, the National Central University of Taiwan, the Space Telescope Science Institute, and the National Aeronautics and Space Administration under Grant No. NNX08AR22G issued through the Planetary Science Division of the NASA Science Mission Directorate, the National Science Foundation Grant No. AST-1238877, the University of Maryland, Eotvos Lorand University (ELTE), and the Los Alamos National Laboratory. (3) The \textit{Wide-field Infrared Survey Explorer}, which is a joint project of the University of California, Los Angeles, and the Jet Propulsion Laboratory/California Institute of Technology, funded by the National Aeronautics and Space Administration. (4) the Two Micron All Sky Survey, which is a joint project of the University of Massachusetts and the Infrared Processing and Analysis Center/California Institute of Technology, funded by the National Aeronautics and Space Administration and the National Science Foundation.  We also used data provided by the SkyView which is developed with generous support from the NASA AISR and ADP programs (P.I. Thomas A. McGlynn) under the auspices of the High Energy Astrophysics Science Archive Research Center (HEASARC) at the NASA/ GSFC Astrophysics Science Division. DKO acknowledges the support of the Department of Atomic Energy, Government of India, under Project Identification No. RTI 4002.
\section*{Data Availability}

The polarimetric data presented in this work are available in this article. The \textit{Gaia} proper motions and distances of the stars are available in \url{https://vizier.u-strasbg.fr/viz-bin/VizieR-3?-source=I/350/gaiaedr3} and \url{https://vizier.u-strasbg.fr/viz-bin/VizieR?-source=I/352}, respectively. The Pan-STARRS \textit{g} and \textit{WISE W}1 data are available in \url{https://vizier.u-strasbg.fr/viz-bin/VizieR?-source=II/349} and \url{https://vizier.u-strasbg.fr/viz-bin/VizieR?-source=II/328}, respectively. The 2MASS data used in this article are available in \url{https://vizier.u-strasbg.fr/viz-bin/VizieR?-source=II/246}.


\bibliographystyle{mnras}
\bibliography{reference}

\begin{thebibliography}{}
\makeatletter
\relax
\def\mn@urlcharsother{\let\do\@makeother \do\$\do\&\do\#\do\^\do\_\do\%\do\~}
\def\mn@doi{\begingroup\mn@urlcharsother \@ifnextchar [ {\mn@doi@}
  {\mn@doi@[]}}
\def\mn@doi@[#1]#2{\def\@tempa{#1}\ifx\@tempa\@empty \href
  {http://dx.doi.org/#2} {doi:#2}\else \href {http://dx.doi.org/#2} {#1}\fi
  \endgroup}
\def\mn@eprint#1#2{\mn@eprint@#1:#2::\@nil}
\def\mn@eprint@arXiv#1{\href {http://arxiv.org/abs/#1} {{\tt arXiv:#1}}}
\def\mn@eprint@dblp#1{\href {http://dblp.uni-trier.de/rec/bibtex/#1.xml}
  {dblp:#1}}
\def\mn@eprint@#1:#2:#3:#4\@nil{\def\@tempa {#1}\def\@tempb {#2}\def\@tempc
  {#3}\ifx \@tempc \@empty \let \@tempc \@tempb \let \@tempb \@tempa \fi \ifx
  \@tempb \@empty \def\@tempb {arXiv}\fi \@ifundefined
  {mn@eprint@\@tempb}{\@tempb:\@tempc}{\expandafter \expandafter \csname
  mn@eprint@\@tempb\endcsname \expandafter{\@tempc}}}

\bibitem[\protect\citeauthoryear{{Alves} \& {Franco}}{{Alves} \&
  {Franco}}{2007}]{2007A&A...470..597A}
{Alves} F.~O.,  {Franco} G.~A.~P.,  2007, \mn@doi [\aap]
  {10.1051/0004-6361:20066759}, \href
  {http://adsabs.harvard.edu/abs/2007A%26A...470..597A} {470, 597}

\bibitem[\protect\citeauthoryear{{Arthur}, {Henney}, {Mellema}, {de Colle}  \&
  {V{\'a}zquez-Semadeni}}{{Arthur} et~al.}{2011}]{2011MNRAS.414.1747A}
{Arthur} S.~J.,  {Henney} W.~J.,  {Mellema} G.,  {de Colle} F.,
  {V{\'a}zquez-Semadeni} E.,  2011, \mn@doi [\mnras]
  {10.1111/j.1365-2966.2011.18507.x}, \href
  {http://adsabs.harvard.edu/abs/2011MNRAS.414.1747A} {414, 1747}

\bibitem[\protect\citeauthoryear{{Bailer-Jones}, {Rybizki}, {Fouesneau},
  {Demleitner}  \& {Andrae}}{{Bailer-Jones} et~al.}{2021}]{2021AJ....161..147B}
{Bailer-Jones} C.~A.~L.,  {Rybizki} J.,  {Fouesneau} M.,  {Demleitner} M.,
  {Andrae} R.,  2021, \mn@doi [\aj] {10.3847/1538-3881/abd806}, \href
  {https://ui.adsabs.harvard.edu/abs/2021AJ....161..147B} {161, 147}

\bibitem[\protect\citeauthoryear{{Baraffe}, {Homeier}, {Allard}  \&
  {Chabrier}}{{Baraffe} et~al.}{2015}]{2015A&A...577A..42B}
{Baraffe} I.,  {Homeier} D.,  {Allard} F.,   {Chabrier} G.,  2015, \mn@doi
  [\aap] {10.1051/0004-6361/201425481}, \href
  {http://adsabs.harvard.edu/abs/2015A%26A...577A..42B} {577, A42}

\bibitem[\protect\citeauthoryear{{Bastien}}{{Bastien}}{1982}]{1982A&AS...48..153B}
{Bastien} P.,  1982, \aaps, \href
  {https://ui.adsabs.harvard.edu/abs/1982A&AS...48..153B} {48, 153}

\bibitem[\protect\citeauthoryear{{Bastien}}{{Bastien}}{1985}]{1985ApJS...59..277B}
{Bastien} P.,  1985, \mn@doi [\apjs] {10.1086/191072}, \href
  {https://ui.adsabs.harvard.edu/abs/1985ApJS...59..277B} {59, 277}

\bibitem[\protect\citeauthoryear{{Bastien} \& {Landstreet}}{{Bastien} \&
  {Landstreet}}{1979}]{1979ApJ...229L.137B}
{Bastien} P.,  {Landstreet} J.~D.,  1979, \mn@doi [\apjl] {10.1086/182947},
  \href {https://ui.adsabs.harvard.edu/abs/1979ApJ...229L.137B} {229, L137}

\bibitem[\protect\citeauthoryear{{Bate}, {Lodato}  \& {Pringle}}{{Bate}
  et~al.}{2010}]{2010MNRAS.401.1505B}
{Bate} M.~R.,  {Lodato} G.,   {Pringle} J.~E.,  2010, \mn@doi [\mnras]
  {10.1111/j.1365-2966.2009.15773.x}, \href
  {https://ui.adsabs.harvard.edu/abs/2010MNRAS.401.1505B} {401, 1505}

\bibitem[\protect\citeauthoryear{{Bayo}, {Barrado}, {Hu{\'e}lamo},
  {Morales-Calder{\'o}n}, {Melo}, {Stauffer}  \& {Stelzer}}{{Bayo}
  et~al.}{2012}]{2012A&A...547A..80B}
{Bayo} A.,  {Barrado} D.,  {Hu{\'e}lamo} N.,  {Morales-Calder{\'o}n} M.,
  {Melo} C.,  {Stauffer} J.,   {Stelzer} B.,  2012, \mn@doi [Astronomy and
  Astrophysics] {10.1051/0004-6361/201219374}, \href
  {https://ui.adsabs.harvard.edu/abs/2012A&A...547A..80B} {547, A80}

\bibitem[\protect\citeauthoryear{{Bertoldi}}{{Bertoldi}}{1989}]{1989ApJ...346..735B}
{Bertoldi} F.,  1989, \mn@doi [\apj] {10.1086/168055}, \href
  {http://adsabs.harvard.edu/abs/1989ApJ...346..735B} {346, 735}

\bibitem[\protect\citeauthoryear{{Bertoldi} \& {McKee}}{{Bertoldi} \&
  {McKee}}{1990}]{1990ApJ...354..529B}
{Bertoldi} F.,  {McKee} C.~F.,  1990, \mn@doi [\apj] {10.1086/168713}, \href
  {http://adsabs.harvard.edu/abs/1990ApJ...354..529B} {354, 529}

\bibitem[\protect\citeauthoryear{{Bisbas}, {W{\"u}nsch}, {Whitworth}, {Hubber}
  \& {Walch}}{{Bisbas} et~al.}{2011}]{2011ApJ...736..142B}
{Bisbas} T.~G.,  {W{\"u}nsch} R.,  {Whitworth} A.~P.,  {Hubber} D.~A.,
  {Walch} S.,  2011, \mn@doi [\apj] {10.1088/0004-637X/736/2/142}, \href
  {https://ui.adsabs.harvard.edu/abs/2011ApJ...736..142B} {736, 142}

\bibitem[\protect\citeauthoryear{{Bressan}, {Marigo}, {Girardi}, {Salasnich},
  {Dal Cero}, {Rubele}  \& {Nanni}}{{Bressan}
  et~al.}{2012}]{2012MNRAS.427..127B}
{Bressan} A.,  {Marigo} P.,  {Girardi} L.,  {Salasnich} B.,  {Dal Cero} C.,
  {Rubele} S.,   {Nanni} A.,  2012, \mn@doi [\mnras]
  {10.1111/j.1365-2966.2012.21948.x}, \href
  {https://ui.adsabs.harvard.edu/abs/2012MNRAS.427..127B} {427, 127}

\bibitem[\protect\citeauthoryear{{Brown} \& {McLean}}{{Brown} \&
  {McLean}}{1977}]{1977A&A....57..141B}
{Brown} J.~C.,  {McLean} I.~S.,  1977, \aap, \href
  {http://adsabs.harvard.edu/abs/1977A%26A....57..141B} {57, 141}

\bibitem[\protect\citeauthoryear{{Chambers} et~al.,}{{Chambers}
  et~al.}{2016}]{2016arXiv161205560C}
{Chambers} K.~C.,  et~al., 2016, preprint, \href
  {http://adsabs.harvard.edu/abs/2016arXiv161205560C} {} (\mn@eprint {arXiv}
  {1612.05560})

\bibitem[\protect\citeauthoryear{{Chauhan}, {Pandey}, {Ogura}, {Ojha}, {Bhatt},
  {Ghosh}  \& {Rawat}}{{Chauhan} et~al.}{2009}]{2009MNRAS.396..964C}
{Chauhan} N.,  {Pandey} A.~K.,  {Ogura} K.,  {Ojha} D.~K.,  {Bhatt} B.~C.,
  {Ghosh} S.~K.,   {Rawat} P.~S.,  2009, \mn@doi [\mnras]
  {10.1111/j.1365-2966.2009.14756.x}, \href
  {https://ui.adsabs.harvard.edu/abs/2009MNRAS.396..964C} {396, 964}

\bibitem[\protect\citeauthoryear{{Chen} \& {Ostriker}}{{Chen} \&
  {Ostriker}}{2014}]{2014ApJ...785...69C}
{Chen} C.-Y.,  {Ostriker} E.~C.,  2014, \mn@doi [\apj]
  {10.1088/0004-637X/785/1/69}, \href
  {https://ui.adsabs.harvard.edu/abs/2014ApJ...785...69C} {785, 69}

\bibitem[\protect\citeauthoryear{{Choudhury}, {Mookerjea}  \&
  {Bhatt}}{{Choudhury} et~al.}{2010}]{2010ApJ...717.1067C}
{Choudhury} R.,  {Mookerjea} B.,   {Bhatt} H.~C.,  2010, \mn@doi [\apj]
  {10.1088/0004-637X/717/2/1067}, \href
  {http://adsabs.harvard.edu/abs/2010ApJ...717.1067C} {717, 1067}

\bibitem[\protect\citeauthoryear{{Crutcher}}{{Crutcher}}{1999}]{1999ApJ...520..706C}
{Crutcher} R.~M.,  1999, \mn@doi [\apj] {10.1086/307483}, \href
  {https://ui.adsabs.harvard.edu/abs/1999ApJ...520..706C} {520, 706}

\bibitem[\protect\citeauthoryear{{Dale}, {Haworth}  \& {Bressert}}{{Dale}
  et~al.}{2015}]{2015MNRAS.450.1199D}
{Dale} J.~E.,  {Haworth} T.~J.,   {Bressert} E.,  2015, \mn@doi [\mnras]
  {10.1093/mnras/stv396}, \href
  {https://ui.adsabs.harvard.edu/abs/2015MNRAS.450.1199D} {450, 1199}

\bibitem[\protect\citeauthoryear{{Dolan} \& {Mathieu}}{{Dolan} \&
  {Mathieu}}{2001}]{2001AJ....121.2124D}
{Dolan} C.~J.,  {Mathieu} R.~D.,  2001, \mn@doi [\aj] {10.1086/319946}, \href
  {https://ui.adsabs.harvard.edu/abs/2001AJ....121.2124D} {121, 2124}

\bibitem[\protect\citeauthoryear{{Elmegreen}}{{Elmegreen}}{1998}]{1998ASPC..148..150E}
{Elmegreen} B.~G.,  1998, in {Woodward} C.~E.,  {Shull} J.~M.,   {Thronson} Jr.
  H.~A.,  eds,  Astronomical Society of the Pacific Conference Series Vol. 148,
  Origins. p.~150 (\mn@eprint {} {astro-ph/9712352})

\bibitem[\protect\citeauthoryear{{Elmegreen}}{{Elmegreen}}{2011}]{2011EAS....51...45E}
{Elmegreen} B.~G.,  2011, in {Charbonnel} C.,  {Montmerle} T.,  eds,  EAS
  Publications Series Vol. 51, EAS Publications Series. pp 45--58 (\mn@eprint
  {arXiv} {1101.3112}), \mn@doi{10.1051/eas/1151004}

\bibitem[\protect\citeauthoryear{{Elsasser} \& {Staude}}{{Elsasser} \&
  {Staude}}{1978}]{1978A&A....70L...3E}
{Elsasser} H.,  {Staude} H.~J.,  1978, \aap, \href
  {http://adsabs.harvard.edu/abs/1978A%26A....70L...3E} {70, L3}

\bibitem[\protect\citeauthoryear{{Eswaraiah}, {Pandey}, {Maheswar}, {Medhi},
  {Pandey}, {Ojha}  \& {Chen}}{{Eswaraiah} et~al.}{2011}]{2011MNRAS.411.1418E}
{Eswaraiah} C.,  {Pandey} A.~K.,  {Maheswar} G.,  {Medhi} B.~J.,  {Pandey}
  J.~C.,  {Ojha} D.~K.,   {Chen} W.~P.,  2011, \mn@doi [\mnras]
  {10.1111/j.1365-2966.2010.17780.x}, \href
  {https://ui.adsabs.harvard.edu/abs/2011MNRAS.411.1418E} {411, 1418}

\bibitem[\protect\citeauthoryear{{Eswaraiah}, {Pandey}, {Maheswar}, {Chen},
  {Ojha}  \& {Chandola}}{{Eswaraiah} et~al.}{2012}]{2012MNRAS.419.2587E}
{Eswaraiah} C.,  {Pandey} A.~K.,  {Maheswar} G.,  {Chen} W.~P.,  {Ojha} D.~K.,
   {Chandola} H.~C.,  2012, \mn@doi [\mnras]
  {10.1111/j.1365-2966.2011.19908.x}, \href
  {https://ui.adsabs.harvard.edu/abs/2012MNRAS.419.2587E} {419, 2587}

\bibitem[\protect\citeauthoryear{{Gaia Collaboration}, {Brown}, {Vallenari},
  {Prusti}, {de Bruijne}, {Babusiaux}  \& {Biermann}}{{Gaia Collaboration}
  et~al.}{2020}]{2020arXiv201201533G}
{Gaia Collaboration} {Brown} A.~G.~A.,  {Vallenari} A.,  {Prusti} T.,  {de
  Bruijne} J.~H.~J.,  {Babusiaux} C.,   {Biermann} M.,  2020, arXiv e-prints,
  \href {https://ui.adsabs.harvard.edu/abs/2020arXiv201201533G} {p.
  arXiv:2012.01533}

\bibitem[\protect\citeauthoryear{{Galli} \& {Shu}}{{Galli} \&
  {Shu}}{1993a}]{1993ApJ...417..220G}
{Galli} D.,  {Shu} F.~H.,  1993a, \mn@doi [\apj] {10.1086/173305}, \href
  {https://ui.adsabs.harvard.edu/abs/1993ApJ...417..220G} {417, 220}

\bibitem[\protect\citeauthoryear{{Galli} \& {Shu}}{{Galli} \&
  {Shu}}{1993b}]{1993ApJ...417..243G}
{Galli} D.,  {Shu} F.~H.,  1993b, \mn@doi [\apj] {10.1086/173306}, \href
  {https://ui.adsabs.harvard.edu/abs/1993ApJ...417..243G} {417, 243}

\bibitem[\protect\citeauthoryear{{Getman}, {Feigelson}, {Garmire}, {Broos}  \&
  {Wang}}{{Getman} et~al.}{2007}]{2007ApJ...654..316G}
{Getman} K.~V.,  {Feigelson} E.~D.,  {Garmire} G.,  {Broos} P.,   {Wang} J.,
  2007, \mn@doi [\apj] {10.1086/509112}, \href
  {http://adsabs.harvard.edu/abs/2007ApJ...654..316G} {654, 316}

\bibitem[\protect\citeauthoryear{{Getman}, {Feigelson}, {Luhman},
  {Sicilia-Aguilar}, {Wang}  \& {Garmire}}{{Getman}
  et~al.}{2009}]{2009ApJ...699.1454G}
{Getman} K.~V.,  {Feigelson} E.~D.,  {Luhman} K.~L.,  {Sicilia-Aguilar} A.,
  {Wang} J.,   {Garmire} G.~P.,  2009, \mn@doi [\apj]
  {10.1088/0004-637X/699/2/1454}, \href
  {http://adsabs.harvard.edu/abs/2009ApJ...699.1454G} {699, 1454}

\bibitem[\protect\citeauthoryear{{Haworth}, {Harries}  \& {Acreman}}{{Haworth}
  et~al.}{2012}]{2012MNRAS.426..203H}
{Haworth} T.~J.,  {Harries} T.~J.,   {Acreman} D.~M.,  2012, \mn@doi [\mnras]
  {10.1111/j.1365-2966.2012.21838.x}, \href
  {http://adsabs.harvard.edu/abs/2012MNRAS.426..203H} {426, 203}

\bibitem[\protect\citeauthoryear{{Hayashi}, {Itoh}  \& {Oasa}}{{Hayashi}
  et~al.}{2012}]{2012PASJ...64...96H}
{Hayashi} M.,  {Itoh} Y.,   {Oasa} Y.,  2012, \mn@doi [\pasj]
  {10.1093/pasj/64.5.96}, \href
  {http://adsabs.harvard.edu/abs/2012PASJ...64...96H} {64, 96}

\bibitem[\protect\citeauthoryear{{Hennebelle} \& {Inutsuka}}{{Hennebelle} \&
  {Inutsuka}}{2019}]{2019FrASS...6....5H}
{Hennebelle} P.,  {Inutsuka} S.-i.,  2019, \mn@doi [Frontiers in Astronomy and
  Space Sciences] {10.3389/fspas.2019.00005}, \href
  {https://ui.adsabs.harvard.edu/abs/2019FrASS...6....5H} {6, 5}

\bibitem[\protect\citeauthoryear{{Henney}, {Arthur}, {de Colle}  \&
  {Mellema}}{{Henney} et~al.}{2009}]{2009MNRAS.398..157H}
{Henney} W.~J.,  {Arthur} S.~J.,  {de Colle} F.,   {Mellema} G.,  2009, \mn@doi
  [\mnras] {10.1111/j.1365-2966.2009.15153.x}, \href
  {http://adsabs.harvard.edu/abs/2009MNRAS.398..157H} {398, 157}

\bibitem[\protect\citeauthoryear{{Hosoya}, {Itoh}, {Oasa}, {Gupta}  \&
  {Sen}}{{Hosoya} et~al.}{2020}]{2020arXiv200600219H}
{Hosoya} K.,  {Itoh} Y.,  {Oasa} Y.,  {Gupta} R.,   {Sen} A.~K.,  2020, arXiv
  e-prints, \href {https://ui.adsabs.harvard.edu/abs/2020arXiv200600219H} {p.
  arXiv:2006.00219}

\bibitem[\protect\citeauthoryear{{Ikeda} et~al.,}{{Ikeda}
  et~al.}{2008}]{2008AJ....135.2323I}
{Ikeda} H.,  et~al., 2008, \mn@doi [\aj] {10.1088/0004-6256/135/6/2323}, \href
  {http://adsabs.harvard.edu/abs/2008AJ....135.2323I} {135, 2323}

\bibitem[\protect\citeauthoryear{{Jones}}{{Jones}}{1997}]{1997MmSAI..68..833J}
{Jones} B.~F.,  1997, \memsai, \href
  {https://ui.adsabs.harvard.edu/abs/1997MmSAI..68..833J} {68, 833}

\bibitem[\protect\citeauthoryear{{Joshi}, {Deshpande}  \&
  {Kulshrestha}}{{Joshi} et~al.}{1987}]{1987IAUS..122..135J}
{Joshi} U.~C.,  {Deshpande} M.~R.,   {Kulshrestha} A.~K.,  1987, in
  {Appenzeller} I.,  {Jordan} C.,  eds,  IAU Symposium Vol. 122, Circumstellar
  Matter. pp 135--136

\bibitem[\protect\citeauthoryear{{Kahn}}{{Kahn}}{1954}]{1954BAN....12..187K}
{Kahn} F.~D.,  1954, \bain, \href
  {https://ui.adsabs.harvard.edu/abs/1954BAN....12..187K} {12, 187}

\bibitem[\protect\citeauthoryear{{Knude} \& {Hog}}{{Knude} \&
  {Hog}}{1998}]{1998A&A...338..897K}
{Knude} J.,  {Hog} E.,  1998, \aap, \href
  {http://adsabs.harvard.edu/abs/1998A%26A...338..897K} {338, 897}

\bibitem[\protect\citeauthoryear{{Koenig}, {Hillenbrand}, {Padgett}  \&
  {DeFelippis}}{{Koenig} et~al.}{2015}]{2015AJ....150..100K}
{Koenig} X.,  {Hillenbrand} L.~A.,  {Padgett} D.~L.,   {DeFelippis} D.,  2015,
  \mn@doi [\aj] {10.1088/0004-6256/150/4/100}, \href
  {https://ui.adsabs.harvard.edu/abs/2015AJ....150..100K} {150, 100}

\bibitem[\protect\citeauthoryear{{Kounkel} et~al.,}{{Kounkel}
  et~al.}{2018}]{2018AJ....156...84K}
{Kounkel} M.,  et~al., 2018, \mn@doi [\aj] {10.3847/1538-3881/aad1f1}, \href
  {https://ui.adsabs.harvard.edu/abs/2018AJ....156...84K} {156, 84}

\bibitem[\protect\citeauthoryear{{Lefloch} \& {Lazareff}}{{Lefloch} \&
  {Lazareff}}{1994}]{1994A&A...289..559L}
{Lefloch} B.,  {Lazareff} B.,  1994, \aap, \href
  {http://adsabs.harvard.edu/abs/1994A%26A...289..559L} {289, 559}

\bibitem[\protect\citeauthoryear{Lindegren}{Lindegren}{2018}]{LL:LL-124}
Lindegren L.,  2018, {R}e-normalising the astrometric chi-square in {G}aia
  {D}{R}2, GAIA-C3-TN-LU-LL-124, \url
  {http://www.rssd.esa.int/doc_fetch.php?id=3757412}

\bibitem[\protect\citeauthoryear{{Lorenzetti} et~al.,}{{Lorenzetti}
  et~al.}{2011}]{2011ApJ...732...69L}
{Lorenzetti} D.,  et~al., 2011, \mn@doi [\apj] {10.1088/0004-637X/732/2/69},
  \href {https://ui.adsabs.harvard.edu/abs/2011ApJ...732...69L} {732, 69}

\bibitem[\protect\citeauthoryear{{Luhman}, {Mamajek}, {Allen}  \&
  {Cruz}}{{Luhman} et~al.}{2009}]{2009ApJ...703..399L}
{Luhman} K.~L.,  {Mamajek} E.~E.,  {Allen} P.~R.,   {Cruz} K.~L.,  2009,
  \mn@doi [\apj] {10.1088/0004-637X/703/1/399}, \href
  {https://ui.adsabs.harvard.edu/abs/2009ApJ...703..399L} {703, 399}

\bibitem[\protect\citeauthoryear{{Mackey} \& {Lim}}{{Mackey} \&
  {Lim}}{2011}]{2011MNRAS.412.2079M}
{Mackey} J.,  {Lim} A.~J.,  2011, \mn@doi [\mnras]
  {10.1111/j.1365-2966.2010.18043.x}, \href
  {http://adsabs.harvard.edu/abs/2011MNRAS.412.2079M} {412, 2079}

\bibitem[\protect\citeauthoryear{{Manset}, {Bastien}, {M{\'e}nard}, {Bertout},
  {Le van Suu}  \& {Boivin}}{{Manset} et~al.}{2009}]{2009A&A...499..137M}
{Manset} N.,  {Bastien} P.,  {M{\'e}nard} F.,  {Bertout} C.,  {Le van Suu} A.,
   {Boivin} L.,  2009, \mn@doi [\aap] {10.1051/0004-6361/200810945}, \href
  {https://ui.adsabs.harvard.edu/abs/2009A&A...499..137M} {499, 137}

\bibitem[\protect\citeauthoryear{{Matsuyanagi}, {Itoh}, {Sugitani}, {Oasa},
  {Mukai}  \& {Tamura}}{{Matsuyanagi} et~al.}{2006}]{2006PASJ...58L..29M}
{Matsuyanagi} I.,  {Itoh} Y.,  {Sugitani} K.,  {Oasa} Y.,  {Mukai} T.,
  {Tamura} M.,  2006, \pasj, \href
  {http://adsabs.harvard.edu/abs/2006PASJ...58L..29M} {58, L29}

\bibitem[\protect\citeauthoryear{{M{\'e}nard}}{{M{\'e}nard}}{2005}]{2005ASPC..343..128M}
{M{\'e}nard} F.~C.,  2005, in {Adamson} A.,  {Aspin} C.,  {Davis} C.,
  {Fujiyoshi} T.,  eds,  Astronomical Society of the Pacific Conference Series
  Vol. 343, Astronomical Polarimetry: Current Status and Future Directions.
  p.~128

\bibitem[\protect\citeauthoryear{{M{\'e}nard} \& {Duch{\^e}ne}}{{M{\'e}nard} \&
  {Duch{\^e}ne}}{2004}]{2004A&A...425..973M}
{M{\'e}nard} F.,  {Duch{\^e}ne} G.,  2004, \mn@doi [\aap]
  {10.1051/0004-6361:20041338}, \href
  {http://adsabs.harvard.edu/abs/2004A%26A...425..973M} {425, 973}

\bibitem[\protect\citeauthoryear{{Miao}, {White}, {Thompson}  \&
  {Nelson}}{{Miao} et~al.}{2009}]{2009ApJ...692..382M}
{Miao} J.,  {White} G.~J.,  {Thompson} M.~A.,   {Nelson} R.~P.,  2009, \mn@doi
  [\apj] {10.1088/0004-637X/692/1/382}, \href
  {https://ui.adsabs.harvard.edu/abs/2009ApJ...692..382M} {692, 382}

\bibitem[\protect\citeauthoryear{{Morgan}, {Thompson}, {Urquhart}, {White}  \&
  {Miao}}{{Morgan} et~al.}{2004}]{2004A&A...426..535M}
{Morgan} L.~K.,  {Thompson} M.~A.,  {Urquhart} J.~S.,  {White} G.~J.,   {Miao}
  J.,  2004, \mn@doi [\aap] {10.1051/0004-6361:20040226}, \href
  {https://ui.adsabs.harvard.edu/abs/2004A&A...426..535M} {426, 535}

\bibitem[\protect\citeauthoryear{{Neha}, {Maheswar}, {Soam}, {Lee}  \&
  {Tej}}{{Neha} et~al.}{2016}]{2016A&A...588A..45N}
{Neha} S.,  {Maheswar} G.,  {Soam} A.,  {Lee} C.~W.,   {Tej} A.,  2016, \mn@doi
  [\aap] {10.1051/0004-6361/201526845}, \href
  {https://ui.adsabs.harvard.edu/abs/2016A&A...588A..45N} {588, A45}

\bibitem[\protect\citeauthoryear{{Neha}, {Maheswar}, {Soam}  \& {Lee}}{{Neha}
  et~al.}{2018}]{2018MNRAS.476.4442N}
{Neha} S.,  {Maheswar} G.,  {Soam} A.,   {Lee} C.~W.,  2018, \mn@doi [\mnras]
  {10.1093/mnras/sty485}, \href
  {https://ui.adsabs.harvard.edu/abs/2018MNRAS.476.4442N} {476, 4442}

\bibitem[\protect\citeauthoryear{{Ogura}, {Sugitani}  \& {Pickles}}{{Ogura}
  et~al.}{2002}]{2002AJ....123.2597O}
{Ogura} K.,  {Sugitani} K.,   {Pickles} A.,  2002, \mn@doi [\aj]
  {10.1086/339976}, \href {http://adsabs.harvard.edu/abs/2002AJ....123.2597O}
  {123, 2597}

\bibitem[\protect\citeauthoryear{{Oort}}{{Oort}}{1954}]{1954BAN....12..177O}
{Oort} J.~H.,  1954, \bain, \href
  {https://ui.adsabs.harvard.edu/abs/1954BAN....12..177O} {12, 177}

\bibitem[\protect\citeauthoryear{{Panwar}, {Chen}, {Pandey}, {Samal}, {Ogura},
  {Ojha}, {Jose}  \& {Bhatt}}{{Panwar} et~al.}{2014}]{2014MNRAS.443.1614P}
{Panwar} N.,  {Chen} W.~P.,  {Pandey} A.~K.,  {Samal} M.~R.,  {Ogura} K.,
  {Ojha} D.~K.,  {Jose} J.,   {Bhatt} B.~C.,  2014, \mn@doi [\mnras]
  {10.1093/mnras/stu1244}, \href
  {http://adsabs.harvard.edu/abs/2014MNRAS.443.1614P} {443, 1614}

\bibitem[\protect\citeauthoryear{{Ramaprakash}, {Gupta}, {Sen}  \&
  {Tandon}}{{Ramaprakash} et~al.}{1998}]{1998A&AS..128..369R}
{Ramaprakash} A.~N.,  {Gupta} R.,  {Sen} A.~K.,   {Tandon} S.~N.,  1998,
  \mn@doi [\aaps] {10.1051/aas:1998148}, \href
  {http://adsabs.harvard.edu/abs/1998A%26AS..128..369R} {128, 369}

\bibitem[\protect\citeauthoryear{{Rautela}, {Joshi}  \& {Pandey}}{{Rautela}
  et~al.}{2004}]{2004BASI...32..159R}
{Rautela} B.~S.,  {Joshi} G.~C.,   {Pandey} J.~C.,  2004, Bulletin of the
  Astronomical Society of India, \href
  {http://adsabs.harvard.edu/abs/2004BASI...32..159R} {32, 159}

\bibitem[\protect\citeauthoryear{{Rebull} et~al.,}{{Rebull}
  et~al.}{2013}]{2013AJ....145...15R}
{Rebull} L.~M.,  et~al., 2013, \mn@doi [\aj] {10.1088/0004-6256/145/1/15},
  \href {https://ui.adsabs.harvard.edu/abs/2013AJ....145...15R} {145, 15}

\bibitem[\protect\citeauthoryear{{Saha}, {Gopinathan}, {Kamath}, {Lee},
  {Puravankara}, {Mathew}  \& {Sharma}}{{Saha}
  et~al.}{2020}]{2020MNRAS.494.5851S}
{Saha} P.,  {Gopinathan} M.,  {Kamath} U.,  {Lee} C.~W.,  {Puravankara} M.,
  {Mathew} B.,   {Sharma} E.,  2020, \mn@doi [\mnras] {10.1093/mnras/staa1053},
  \href {https://ui.adsabs.harvard.edu/abs/2020MNRAS.494.5851S} {494, 5851}

\bibitem[\protect\citeauthoryear{{Saha}, {Maheswar}, {Mathew}  \&
  {Kamath}}{{Saha} et~al.}{2021}]{2021arXiv210412061S}
{Saha} P.,  {Maheswar} G.,  {Mathew} B.,   {Kamath} U.~S.,  2021, arXiv
  e-prints, \href {https://ui.adsabs.harvard.edu/abs/2021arXiv210412061S} {p.
  arXiv:2104.12061}

\bibitem[\protect\citeauthoryear{{Sato}}{{Sato}}{1988}]{1988PThPS..96...37S}
{Sato} S.,  1988, \mn@doi [Progress of Theoretical Physics Supplement]
  {10.1143/PTPS.96.37}, \href
  {https://ui.adsabs.harvard.edu/abs/1988PThPS..96...37S} {96, 37}

\bibitem[\protect\citeauthoryear{{Schmidt}, {Elston}  \& {Lupie}}{{Schmidt}
  et~al.}{1992}]{1992AJ....104.1563S}
{Schmidt} G.~D.,  {Elston} R.,   {Lupie} O.~L.,  1992, \mn@doi [\aj]
  {10.1086/116341}, \href {http://adsabs.harvard.edu/abs/1992AJ....104.1563S}
  {104, 1563}

\bibitem[\protect\citeauthoryear{{Sharma} et~al.,}{{Sharma}
  et~al.}{2016}]{2016AJ....151..126S}
{Sharma} S.,  et~al., 2016, \mn@doi [\aj] {10.3847/0004-6256/151/5/126}, \href
  {http://adsabs.harvard.edu/abs/2016AJ....151..126S} {151, 126}

\bibitem[\protect\citeauthoryear{{Skrutskie} et~al.,}{{Skrutskie}
  et~al.}{2006}]{2006AJ....131.1163S}
{Skrutskie} M.~F.,  et~al., 2006, \mn@doi [\aj] {10.1086/498708}, \href
  {https://ui.adsabs.harvard.edu/abs/2006AJ....131.1163S} {131, 1163}

\bibitem[\protect\citeauthoryear{{Soam}, {Maheswar}, {Bhatt}, {Lee}  \&
  {Ramaprakash}}{{Soam} et~al.}{2013}]{2013MNRAS.432.1502S}
{Soam} A.,  {Maheswar} G.,  {Bhatt} H.~C.,  {Lee} C.~W.,   {Ramaprakash} A.~N.,
   2013, \mn@doi [\mnras] {10.1093/mnras/stt576}, \href
  {https://ui.adsabs.harvard.edu/abs/2013MNRAS.432.1502S} {432, 1502}

\bibitem[\protect\citeauthoryear{{Soam}, {Maheswar}, {Lee}, {Dib}, {Bhatt},
  {Tamura}  \& {Kim}}{{Soam} et~al.}{2015}]{2015A&A...573A..34S}
{Soam} A.,  {Maheswar} G.,  {Lee} C.~W.,  {Dib} S.,  {Bhatt} H.~C.,  {Tamura}
  M.,   {Kim} G.,  2015, \mn@doi [\aap] {10.1051/0004-6361/201322536}, \href
  {https://ui.adsabs.harvard.edu/abs/2015A&A...573A..34S} {573, A34}

\bibitem[\protect\citeauthoryear{{Soam}, {Maheswar}, {Lee}, {Neha}  \&
  {Andersson}}{{Soam} et~al.}{2017}]{2017MNRAS.465..559S}
{Soam} A.,  {Maheswar} G.,  {Lee} C.~W.,  {Neha} S.,   {Andersson} B.~G.,
  2017, \mn@doi [\mnras] {10.1093/mnras/stw2649}, \href
  {https://ui.adsabs.harvard.edu/abs/2017MNRAS.465..559S} {465, 559}

\bibitem[\protect\citeauthoryear{{Straizys}, {Cernis}, {Kazlauskas}  \&
  {Meistas}}{{Straizys} et~al.}{1992}]{1992BaltA...1..149S}
{Straizys} V.,  {Cernis} K.,  {Kazlauskas} A.,   {Meistas} E.,  1992, \mn@doi
  [Baltic Astronomy] {10.1515/astro-1992-0203}, \href
  {http://adsabs.harvard.edu/abs/1992BaltA...1..149S} {1, 149}

\bibitem[\protect\citeauthoryear{{Strom}}{{Strom}}{1977}]{1977IAUS...75..179S}
{Strom} S.~E.,  1977, in {de Jong} T.,  {Maeder} A.,   {Pikel'Ner} S.~B.,  eds,
   IAU Symposium Vol. 75, Star Formation. p.~179

\bibitem[\protect\citeauthoryear{{Sugitani} \& {Ogura}}{{Sugitani} \&
  {Ogura}}{1994}]{1994ApJS...92..163S}
{Sugitani} K.,  {Ogura} K.,  1994, \mn@doi [\apjs] {10.1086/191964}, \href
  {http://adsabs.harvard.edu/abs/1994ApJS...92..163S} {92, 163}

\bibitem[\protect\citeauthoryear{{Sugitani}, {Fukui}  \& {Ogura}}{{Sugitani}
  et~al.}{1991}]{1991ApJS...77...59S}
{Sugitani} K.,  {Fukui} Y.,   {Ogura} K.,  1991, \mn@doi [\apjs]
  {10.1086/191597}, \href {http://adsabs.harvard.edu/abs/1991ApJS...77...59S}
  {77, 59}

\bibitem[\protect\citeauthoryear{{Sugitani}, {Tamura}  \& {Ogura}}{{Sugitani}
  et~al.}{1995}]{1995ApJ...455L..39S}
{Sugitani} K.,  {Tamura} M.,   {Ogura} K.,  1995, \mn@doi [\apjl]
  {10.1086/309808}, \href {http://adsabs.harvard.edu/abs/1995ApJ...455L..39S}
  {455, L39}

\bibitem[\protect\citeauthoryear{{Thompson}, {Urquhart}  \& {White}}{{Thompson}
  et~al.}{2004}]{2004A&A...415..627T}
{Thompson} M.~A.,  {Urquhart} J.~S.,   {White} G.~J.,  2004, \mn@doi [\aap]
  {10.1051/0004-6361:20031681}, \href
  {https://ui.adsabs.harvard.edu/abs/2004A&A...415..627T} {415, 627}

\bibitem[\protect\citeauthoryear{{Vaillancourt}}{{Vaillancourt}}{2006}]{2006PASP..118.1340V}
{Vaillancourt} J.~E.,  2006, \mn@doi [\pasp] {10.1086/507472}, \href
  {https://ui.adsabs.harvard.edu/abs/2006PASP..118.1340V} {118, 1340}

\bibitem[\protect\citeauthoryear{{Vincke} \& {Pfalzner}}{{Vincke} \&
  {Pfalzner}}{2016}]{2016ApJ...828...48V}
{Vincke} K.,  {Pfalzner} S.,  2016, \mn@doi [\apj]
  {10.3847/0004-637X/828/1/48}, \href
  {https://ui.adsabs.harvard.edu/abs/2016ApJ...828...48V} {828, 48}

\bibitem[\protect\citeauthoryear{{Vink}, {Drew}, {Harries}, {Oudmaijer}  \&
  {Unruh}}{{Vink} et~al.}{2005}]{2005MNRAS.359.1049V}
{Vink} J.~S.,  {Drew} J.~E.,  {Harries} T.~J.,  {Oudmaijer} R.~D.,   {Unruh}
  Y.,  2005, \mn@doi [\mnras] {10.1111/j.1365-2966.2005.08969.x}, \href
  {https://ui.adsabs.harvard.edu/abs/2005MNRAS.359.1049V} {359, 1049}

\bibitem[\protect\citeauthoryear{{Wardle} \& {Kronberg}}{{Wardle} \&
  {Kronberg}}{1974}]{1974ApJ...194..249W}
{Wardle} J.~F.~C.,  {Kronberg} P.~P.,  1974, \mn@doi [\apj] {10.1086/153240},
  \href {https://ui.adsabs.harvard.edu/abs/1974ApJ...194..249W} {194, 249}

\bibitem[\protect\citeauthoryear{{Whittet}, {Prusti}, {Franco}, {Gerakines},
  {Kilkenny}, {Larson}  \& {Wesselius}}{{Whittet}
  et~al.}{1997}]{1997A&A...327.1194W}
{Whittet} D.~C.~B.,  {Prusti} T.,  {Franco} G.~A.~P.,  {Gerakines} P.~A.,
  {Kilkenny} D.,  {Larson} K.~A.,   {Wesselius} P.~R.,  1997, \aap, \href
  {http://adsabs.harvard.edu/abs/1997A%26A...327.1194W} {327, 1194}

\bibitem[\protect\citeauthoryear{{Wilking}, {Vrba}  \& {Sullivan}}{{Wilking}
  et~al.}{2015}]{2015ApJ...815....2W}
{Wilking} B.~A.,  {Vrba} F.~J.,   {Sullivan} T.,  2015, \mn@doi [\apj]
  {10.1088/0004-637X/815/1/2}, \href
  {https://ui.adsabs.harvard.edu/abs/2015ApJ...815....2W} {815, 2}

\bibitem[\protect\citeauthoryear{{Wright} et~al.,}{{Wright}
  et~al.}{2010}]{2010AJ....140.1868W}
{Wright} E.~L.,  et~al., 2010, \mn@doi [\aj] {10.1088/0004-6256/140/6/1868},
  \href {https://ui.adsabs.harvard.edu/abs/2010AJ....140.1868W} {140, 1868}

\bibitem[\protect\citeauthoryear{{Yang}, {Li}, {Looney}  \& {Stephens}}{{Yang}
  et~al.}{2016}]{2016MNRAS.456.2794Y}
{Yang} H.,  {Li} Z.-Y.,  {Looney} L.,   {Stephens} I.,  2016, \mn@doi [\mnras]
  {10.1093/mnras/stv2633}, \href
  {https://ui.adsabs.harvard.edu/abs/2016MNRAS.456.2794Y} {456, 2794}

\bibitem[\protect\citeauthoryear{{Zari}, {Hashemi}, {Brown}, {Jardine}  \& {de
  Zeeuw}}{{Zari} et~al.}{2018}]{2018A&A...620A.172Z}
{Zari} E.,  {Hashemi} H.,  {Brown} A.~G.~A.,  {Jardine} K.,   {de Zeeuw} P.~T.,
   2018, \mn@doi [\aap] {10.1051/0004-6361/201834150}, \href
  {https://ui.adsabs.harvard.edu/abs/2018A&A...620A.172Z} {620, A172}

\makeatother
\end{thebibliography}



\appendix
\section{tables}
\begin{table*}
	\begin{center}
		\caption{Properties of the previously known candidate YSOs towards BRC 18 from \textit{Gaia} EDR3.}
		\label{tab:yso_gaia}
		\renewcommand{\arraystretch}{1.3}
		\scriptsize
		\begin{tabular}{lccccccccccc} 
			\hline
			\#& Source ID& RA(2016)  & Dec(2016)  & $\Delta$RA & $\Delta$Dec & $d\pm\Delta d$ & $\mu_{\alpha\star}\pm\Delta\mu_{\alpha\star}$ & $\mu_{\delta}\pm\Delta\mu_{\delta}$ & $G\pm\Delta G$ & $G_\mathrm{BP}\pm\Delta G_\mathrm{BP}$ & $G_\mathrm{RP}\pm\Delta G_\mathrm{RP}$\\
			& (\textit{Gaia} EDR3) & ($^{\circ}$) & ($^{\circ}$) & ($^{\prime\prime}$) & ($^{\prime\prime}$) & (pc)& (mas yr$^{-1}$) & (mas yr$^{-1}$) & (mag)&(mag)&(mag)\\
			(1)&(2)&(3)&(4)&(5)&(6)&(7)&(8)&(9)&(10)&(11)&(12)\\
			\hline
1  & 3336168314489240704 & 85.660435 & 9.310332 & 6658.4 &-2245.7&397$_{-4}^{4}$& 2.054$\pm$0.026 & -2.187$\pm$0.018 & 14.580$\pm$0.003 & 15.405$\pm$0.009 & 13.665$\pm$0.007\\
2  & 3336168043908536704 & 85.680581 & 9.287969 & 6730.1 &-2326.3&383$_{-9}^{8}$& 1.409$\pm$0.070 & -2.557$\pm$0.047 & 16.419$\pm$0.003 & 18.436$\pm$0.018 & 15.080$\pm$0.005\\
3  & 3336155189070384384 & 85.729819 & 9.246989 & 6905.3 &-2473.8&390$_{-6}^{6}$& 2.076$\pm$0.054 & -2.141$\pm$0.037 & 15.995$\pm$0.003 & 17.431$\pm$0.008 & 14.818$\pm$0.005\\		
4  & 3336152375867856512 & 85.747103 & 9.164555 & 6967.5 &-2770.5&390$_{-4}^{6}$& 1.506$\pm$0.035 & -2.735$\pm$0.025 & 15.059$\pm$0.003 & 16.172$\pm$0.012 & 13.977$\pm$0.005\\			
5  & 3336161549917961088 & 85.802402 & 9.283812 & 7162.5 &-2341.2&394$_{-4}^{5}$& 2.011$\pm$0.036 & -2.482$\pm$0.023 & 15.179$\pm$0.005 & 16.392$\pm$0.020 & 14.070$\pm$0.012\\	
6  & 3336180585212965376 & 85.814037 & 9.491866 & 7201.6 &-1592.2&407$_{-4}^{5}$ & 2.137$\pm$0.027 & -2.488$\pm$0.018 & 14.675$\pm$0.003 & 15.555$\pm$0.021 & 13.638$\pm$0.008\\
7  & 3336160690924950272 & 85.873411 & 9.283380 & 7414.6 &-2342.8&400$_{-9}^{10}$& 2.180$\pm$0.077 & -2.473$\pm$0.046 & 16.390$\pm$0.003 & 17.857$\pm$0.015 & 15.201$\pm$0.005\\
8  & 3336177454181082496 & 85.890385 & 9.496435 & 7472.5 &-1575.8&401$_{-15}^{13}$& 2.106$\pm$0.091 & -2.722$\pm$0.065 & 17.316$\pm$0.003 & 19.431$\pm$0.025 & 15.958$\pm$0.005\\
9  & 3336156121079561344 & 85.914989 & 9.178026 & 7563.3 &-2722.0&370$_{-9}^{15}$& 2.170$\pm$0.103 & -2.299$\pm$0.065 & 16.676$\pm$0.003 & 18.491$\pm$0.017 & 15.345$\pm$0.005\\
10 & 3336149867606924288 & 85.935460 & 9.133633 & 7636.5 &-2881.9&429$_{-32}^{30}$ & 1.069$\pm$0.183 & -2.405$\pm$0.116 & 18.321$\pm$0.010 & 19.229$\pm$0.048 & 17.202$\pm$0.031\\
11 & 3336156292877976064 & 85.961132 & 9.206919 & 7726.8 &-2618.0&405$_{-9}^{11}$& 2.422$\pm$0.065 & -2.415$\pm$0.040 & 16.295$\pm$0.010 & 17.859$\pm$0.040 & 15.057$\pm$0.024\\
12 & 3336157422453353600 & 85.963992 & 9.227901 & 7736.7 &-2542.5&399$_{-31}^{30}$& 2.189$\pm$0.219 & -2.004$\pm$0.145 & 18.391$\pm$0.003 & 20.377$\pm$0.074 & 16.975$\pm$0.010\\
13 & 3336159419614147584 & 85.968111 & 9.283435 & 7750.7 &-2342.6&394$_{-7}^{6}$& 2.690$\pm$0.046 & -2.470$\pm$0.031 & 15.716$\pm$0.003 & 17.415$\pm$0.006 & 14.464$\pm$0.004\\
14 & 3336157873425924992 & 85.968650 & 9.266585 & 7752.8 &-2403.2&384$_{-15}^{13}$& 2.227$\pm$0.107 & -2.585$\pm$0.074 & 17.422$\pm$0.003 & 19.441$\pm$0.035 & 16.078$\pm$0.006\\
15 & 3336102897844569856 & 85.972911 & 9.119901 & 7769.6 &-2931.3&394$_{-9}^{8}$& 1.414$\pm$0.066 & -2.423$\pm$0.043 & 16.228$\pm$0.009 & 17.575$\pm$0.043 & 15.000$\pm$0.025\\	
16 & 3336159557053093248 & 85.985790 & 9.304671 & 7813.2 &-2266.1&371$_{-9}^{9}$& 2.388$\pm$0.062 & -2.181$\pm$0.044 & 16.397$\pm$0.016 & 17.831$\pm$0.051 & 15.204$\pm$0.043\\
17 & 3336159282175185792 & 86.006802 & 9.289093 & 7888.0 &-2322.2&393$_{-4}^{5}$& 2.753$\pm$0.028 & -2.600$\pm$0.019 & 14.750$\pm$0.003 & 15.793$\pm$0.008 & 13.726$\pm$0.005\\		
18 & 3336109048235617536 & 86.041829 & 9.144624 & 8014.0 &-2842.3&414$_{-13}^{12}$& 1.884$\pm$0.090 & -2.221$\pm$0.052 & 16.814$\pm$0.008 & 17.926$\pm$0.035 & 15.703$\pm$0.015\\
19 & 3336107429033839360 & 86.053240 & 9.086216 & 8055.2 &-3052.6&391$_{-8}^{9}$& 1.496$\pm$0.075 & -2.510$\pm$0.046 & 16.625$\pm$0.004 & 18.211$\pm$0.013 & 15.350$\pm$0.014\\			
20 & 3336108223604259456 & 86.061986 & 9.135564 & 8085.7 &-2874.9&385$_{-8}^{7}$& 1.588$\pm$0.050 & -2.323$\pm$0.031 & 15.728$\pm$0.003 & 17.242$\pm$0.009 & 14.524$\pm$0.007\\
21 & 3336111006743055360 & 86.079400 & 9.209260 & 8146.6 &-2609.6&382$_{-12}^{7}$ & 2.229$\pm$0.078 & -2.416$\pm$0.048 & 16.458$\pm$0.003 & 18.186$\pm$0.023 & 15.024$\pm$0.005\\
22 & 3336158594980405376 & 86.080847 & 9.271995 & 8151.0 &-2383.8&409$_{-9}^{12}$& 1.613$\pm$0.056 & -2.534$\pm$0.036 & 16.170$\pm$0.003 & 17.742$\pm$0.009 & 14.948$\pm$0.005\\
23 & 3336158521965029760 & 86.087587 & 9.267106 & 8175.0 &-2401.4&382$_{-7}^{8}$& 2.546$\pm$0.061 & -2.198$\pm$0.039 & 16.189$\pm$0.003 & 17.999$\pm$0.009 & 14.910$\pm$0.004\\			
24 & 3336108361043207040 & 86.090606 & 9.148215 & 8187.2 &-2829.4&399$_{-8}^{9}$& 1.699$\pm$0.061 & -2.753$\pm$0.039 & 15.938$\pm$0.003 & 17.824$\pm$0.022 & 14.462$\pm$0.007\\
25 & 3336109937294863744 & 86.095395 & 9.176425 & 8203.8 &-2727.8&394$_{-5}^{6}$& 2.635$\pm$0.035 & -2.412$\pm$0.022 & 15.059$\pm$0.003 & 16.228$\pm$0.006 & 13.978$\pm$0.005\\
26 & 3336110834944361984 & 86.107235 & 9.200320 & 8245.6 &-2641.8&387$_{-8}^{7}$& 3.095$\pm$0.057 & -2.596$\pm$0.036 & 16.160$\pm$0.003 & 17.781$\pm$0.015 & 14.913$\pm$0.007\\
27 & 3336109902935129088 & 86.112112 & 9.178385 & 8263.1 &-2720.8&393$_{-6}^{8}$ & 2.891$\pm$0.056 & -2.444$\pm$0.035 & 16.221$\pm$0.003 & 17.718$\pm$0.007 & 15.033$\pm$0.004\\
28 & 3336104955132704640 & 86.119424 & 9.105122 & 8290.0 &-2984.5&416$_{-22}^{27}$& 2.933$\pm$0.150 & -2.390$\pm$0.097 & 17.918$\pm$0.003 & 20.044$\pm$0.062 & 16.563$\pm$0.007\\
29 & 3336256176637551232 & 86.137795 & 9.446547 & 8351.0 &-1755.4&383$_{-4}^{3}$& 2.782$\pm$0.029 & -1.843$\pm$0.019 & 14.865$\pm$0.003 & 16.315$\pm$0.004 & 13.683$\pm$0.004\\
30 & 3336110525706709120 & 86.154261 & 9.222194 & 8412.2 &-2563.0&384$_{-6}^{7}$& 2.709$\pm$0.043 & -2.393$\pm$0.027 & 15.580$\pm$0.003 & 17.301$\pm$0.008 & 14.309$\pm$0.005\\
31 & 3336104787630415616 & 86.157234 & 9.099283 & 8424.3 &-3005.5&409$_{-9}^{11}$& 2.625$\pm$0.081 & -2.558$\pm$0.053 & 16.663$\pm$0.003 & 18.792$\pm$0.018 & 15.309$\pm$0.004\\
32 & 3336256103621173504 & 86.158848 & 9.432716 & 8425.9 &-1805.2&420$_{-13}^{16}$& 1.832$\pm$0.081 & -2.203$\pm$0.056 & 17.188$\pm$0.003 & 18.984$\pm$0.027 & 15.912$\pm$0.005\\
33 & 3336107227171825664 & 86.194115 & 9.189299 & 8554.1 &-2681.5&402$_{-6}^{6}$& 3.053$\pm$0.041 & -2.504$\pm$0.027 & 15.667$\pm$0.003 & 16.959$\pm$0.008 & 14.528$\pm$0.005\\
34 & 3335326299037613056 & 86.226617 & 8.713875 & 8675.5 &-4393.0&417$_{-13}^{14}$& 2.009$\pm$0.088 & -2.621$\pm$0.059 & 17.223$\pm$0.017 & 19.321$\pm$0.081 & 15.809$\pm$0.046\\
    \hline
\end{tabular}\\
\end{center}
\end{table*}
\begin{table*}
    \begin{center}
    \caption{Properties of the newly found comoving sources identified towards BRC 18 from \textit{Gaia} EDR3.}
    \label{tab:comoving_gaia}
    \renewcommand{\arraystretch}{1.3}
    \scriptsize
    \begin{tabular}{lccccccccccc} 
    \hline
    \#& Source ID& RA(2016)  & Dec(2016)  & $\Delta$RA& $\Delta$Dec&$d\pm\Delta d$ & $\mu_{\alpha\star}\pm\Delta\mu_{\alpha\star}$ & $\mu_{\delta}\pm\Delta\mu_{\delta}$ & $G\pm\Delta G$ & $G_\mathrm{BP}\pm\Delta G_\mathrm{BP}$ & $G_\mathrm{RP}\pm\Delta G_\mathrm{RP}$\\
     & (\textit{Gaia} EDR3) & ($^{\circ}$) & ($^{\circ}$) & ($^{\prime\prime}$) & ($^{\prime\prime}$) &(pc)& (mas yr$^{-1}$) & (mas yr$^{-1}$) & (mag)&(mag)&(mag)\\
     (1)&(2)&(3)&(4)&(5)&(6)&(7)&(8)&(9)&(10)&(11)&(12)\\
    \hline
1  & 3336166837021667584 & 85.679980 & 9.228623 & 6728.5 &-2539.9 &407$_{-27}^{21}$ & 1.911$\pm$0.161 & -2.499$\pm$0.108 & 17.851$\pm$0.003 & 20.035$\pm$0.072 & 16.481$\pm$0.006\\
2  & 3336149455287469824 & 85.846633 & 9.091804 & 7321.6 &-3032.4 &405$_{-14}^{20}$ & 2.568$\pm$0.113 & -2.309$\pm$0.079 & 16.151$\pm$0.003&-&-\\
3  & 3336156048063746304 & 85.940706 & 9.171731 & 7654.7 &-2744.7 &400$_{-17}^{29}$ & 2.128$\pm$0.123 & -2.704$\pm$0.085 & 17.312$\pm$0.003 & 19.461$\pm$0.034 & 15.921$\pm$0.005\\
4  & 3336111071166289152 & 86.069824 & 9.222508 & 8112.5 &-2561.9 &371$_{-12}^{11}$ & 2.510$\pm$0.098 & -2.494$\pm$0.064 & 17.234$\pm$0.004 & 18.898$\pm$0.046 & 15.751$\pm$0.006\\    
5  & 3336104134795404928 & 86.093515 & 9.044528 & 8198.7 &-3202.6 &401$_{-19}^{19}$ & 2.066$\pm$0.143 & -2.546$\pm$0.088 & 17.408$\pm$0.003 & 19.350$\pm$0.034 & 16.072$\pm$0.005\\
6  & 3336108391106615168 & 86.102807 & 9.163186 & 8230.3 &-2775.5 &403$_{-27}^{19}$ & 2.356$\pm$0.147 & -2.626$\pm$0.093 & 17.930$\pm$0.003 & 19.966$\pm$0.042 & 16.601$\pm$0.007\\
7  & 3336252564567949824 & 86.151543 & 9.308982 & 8401.5 &-2250.6 &389$_{-21}^{25}$ & 2.636$\pm$0.163 & -2.223$\pm$0.106 & 18.262$\pm$0.003 & 20.685$\pm$0.083 & 16.846$\pm$0.007\\
8  & 3336205663525068800 & 86.181870 & 9.299059 & 8509.3 &-2286.3 &385$_{-22}^{27}$ & 3.195$\pm$0.199 & -2.558$\pm$0.132 &  18.311$\pm$0.003 & 20.480$\pm$0.070 & 16.893$\pm$0.007\\
9  & 3336110182109326208 & 86.182331 & 9.192329 & 8512.2 &-2670.6 &386$_{-17}^{15}$ & 2.511$\pm$0.135 & -2.360$\pm$0.080 & 17.502$\pm$0.003 & 19.949$\pm$0.048 & 16.067$\pm$0.005\\
10 & 3336107364610774272 & 86.219055 & 9.209368 & 8642.4 &-2609.2 &375$_{-8}^{8}$ & 3.337$\pm$0.070 & -2.662$\pm$0.045 & 16.628$\pm$0.003 & 18.785$\pm$0.018 & 15.263$\pm$0.005\\
11 & 3336092830441451008 & 86.330422 & 9.056243 & 9039.8 &-3160.5 &389$_{-15}^{23}$ & 2.802$\pm$0.145 & -2.712$\pm$0.076 &  17.316$\pm$0.003 & 19.503$\pm$0.031 & 15.971$\pm$0.005\\
12 & 3336092864798790144 & 86.343198 & 9.070070 & 9084.9 &-3110.7 &402$_{-4}^{3}$ & 2.441$\pm$0.020 & -2.507$\pm$0.012 & 9.266$\pm$0.003 & 10.029$\pm$0.004 & 8.360$\pm$0.005\\
    \hline
    \end{tabular}\\
    \end{center}
\end{table*}


\bsp	
\label{lastpage}
\end{document}